\documentclass[aps,onecolumn,nofootinbib,floatfix,letterpaper,showpacs,notitlepage]{revtex4-1}
\pdfoutput=1

\usepackage{natbib}
\usepackage{times}
\usepackage{graphicx}
\usepackage{epsfig}
\usepackage{subfigure}
\usepackage{amssymb,amsmath}
\usepackage{hyperref}
\usepackage{setspace}
\usepackage{url}

\hypersetup{
    colorlinks = true,
    linkcolor = blue,
    anchorcolor = blue,
    citecolor = blue,
    filecolor = blue,
    pagecolor = blue,
    urlcolor = blue
}

\begin{document}

\title {$t-b-\tau$ Yukawa unification for $\mu < 0$ with a sub-TeV sparticle spectrum }

\author{Ilia Gogoladze\footnote{
Email: \textcolor{blue}{ilia@bartol.udel.edu}. On leave of absence from:
Andronikashvili Institute of Physics, GAS, Tbilisi, Georgia.}}
\affiliation{Bartol Research Institute, Department of Physics and Astronomy\\
University of Delaware, Newark, Delaware 19716}

\author{Rizwan Khalid\footnote{
Email: \textcolor{blue}{rizwan.hep@gmail.com}. On study leave from:
Centre for Advanced Mathematics \& Physics of the National
University of Sciences \& Technology, H-12, Islamabad, Pakistan. }}
\affiliation{Bartol Research Institute, Department of Physics and Astronomy\\
University of Delaware, Newark, Delaware 19716}

\author{Shabbar Raza\footnote{
Email: \textcolor{blue}{shabbar@udel.edu}. On study leave from:
Department of Physics, FUUAST, Islamabad, Pakistan.}}
\affiliation{Bartol Research Institute, Department of Physics and Astronomy\\
University of Delaware, Newark, Delaware 19716}

\author{Qaisar Shafi}
\affiliation{Bartol Research Institute, Department of Physics and Astronomy\\
University of Delaware, Newark, Delaware 19716}


\begin{abstract}

We show compatibility with all known experimental
constraints of $t-b-\tau$ Yukawa coupling unification in
supersymmetric $SU(4)_c \times SU(2)_L \times SU(2)_R$ which has
non-universal gaugino masses and the MSSM parameter $\mu < 0$. In particular,
the relic neutralino abundance satisfies the WMAP bounds and $\Delta (g-2)_\mu$ is in good
agreement with the observations. We identify benchmark points for the sparticle spectra
which can be tested at the LHC, including those associated with gluino and stau
coannihilation channels, mixed bino-Higgsino state and the $A$-funnel
region. We also briefly discuss prospects for testing Yukawa unification
with the ongoing and planned direct detection experiments.

\end{abstract}

\maketitle

\section{Introduction}

Supersymmetric $SO(10)$, in contrast to its non-supersymmetric version,
yields third family ($t-b-\tau$) Yukawa unification via the unique
renormalizable Yukawa coupling $16  \cdot 16 \cdot 10$, where the
10-plet is assumed to contain the two minimal supersymmteric standard
model (MSSM) Higgs doublets $H_u$ and $H_d$. The 16-plet contains
the 15 chiral fermions per family of the standard model (SM) as
well as right handed neutrino. The implications of this unification
have been extensively explored over the years
\cite{big-422,bigger-422}. More recently, it has been argued in
\cite{Baer:2008jn, Gogoladze:2009ug} that $SO(10)$ Yukawa unification
predicts relatively light ($\leq$ TeV) gluinos, which can be readily
tested \cite{Baer:2009ff} at the Large Hadron Collider (LHC). The
squarks and sleptons  turn out to have masses in the multi-TeV range.
Moreover, it is argued in \cite{Baer:2008jn, Gogoladze:2009ug} that
the lightest neutralino is not a viable cold dark matter candidate,
at least in the simplest models of $SO(10)$ Yukawa unification.

Spurred by these developments we have investigated $t-b-\tau$ Yukawa
unification \cite{Gogoladze:2009ug, Gogoladze:2009bn} in the
framework of supersymmetric $SU(4)_c \times SU(2)_L \times SU(2)_R$ \cite{pati}
(4-2-2, for short), with a positive  supersymmetric bilinear
Higgs parameter ($\mu>0$). The 4-2-2 structure allows us to
consider non-universal gaugino masses while retaining Yukawa
unification. In particular, assuming left-right symmetry \cite{pati,lr},
or more precisely C-parity \cite{c-parity}, we have only one
additional free parameter in the soft supersymmetry breaking (SSB)
sector compared to the $SO(10)$ model. In particular, this allow us
to distinguish $M_2$ from $M_3$, where $M_2$ ($M_3$) denotes the
asymptotic $SU(2)_L$ ($SU(3)_c$) gaugino mass. An important conclusion
reached in \cite{Gogoladze:2009ug, Gogoladze:2009bn} is that with
gaugino non-universality, Yukawa unification in 4-2-2 is compatible
with neutralino dark matter, with gluino co-annihilation
\cite{Profumo:2004wk, Gogoladze:2009ug, Gogoladze:2009bn}
playing an important role.

The main purpose of this paper to extend the 4-2-2 discussion to
the case of $\mu < 0$, where $\mu$ denotes the supersymmetric Higgs
mass parameter. Most authors normally do not consider $\mu < 0$ because
it can create serious disagreement with the measured value of the muon
anomalous magnetic moment $(g-2)_\mu$~\cite{Bennett:2006fi}.
The new contribution to $(g-2)_\mu$ from supersymmetric particles
is proportional to $ \mu M_2 \tan\beta / \tilde{m}^4$ \cite{Stockinger:2006zn},
where $\tilde{m}$ is the heaviest sparticle mass in the loop.
In $SO(10)$ Yukawa unification, the scalar masses are very heavy (multi-TeV),
so that the supersymmetry (SUSY)
contribution to $(g-2)_\mu$ effectively decouples and one is left
with the standard model result \cite{Baer:2008jn, Gogoladze:2009ug}.
On the contrary, in 4-2-2, with both $\mu < 0$ and $M_2<0$,
the SUSY contributions to $(g-2)_\mu$ turn out to be important,
as we will see, and thereby yield significantly improved agreement with
experimental data.

The outline for the rest of the paper is as follows. In Section
\ref{model} we  briefly describe the model and the boundary conditions
for SSB parameters which we employ for our scan. In Section
\ref{constraintsSection} we summarize the scanning procedure and the
experimental constraints that we have employed. In Section
\ref{muneg} we discuss how $\mu<0$ leads to better Yukawa
unification than $\mu>0$. In Section \ref{results}
we present the results from our scan and highlight
some of the predictions of the 4-2-2 model.
The correlation between the spin-independent and
spin-dependent direct detection of dark matter and Yukawa
unification condition is presented in Section \ref{dark} where we also
display some benchmark points. Our conclusions are summarized in
Section \ref{conclusions}.

\section{The 4-2-2 model \label{model}}

In 4-2-2 the 16-plet of $SO(10)$ matter fields consists of
$\psi$ (4, 2, 1) and $\psi_c$ $(\bar{4}, 1, 2)$. The third family Yukawa coupling
$\psi_c \psi H$, where H(1,2,2) denotes the bi-doublet (1,2,2),  yields
the following relation valid at $M_{\rm GUT}$,
\begin{align}
Y_t = Y_b = Y_{\tau} = Y_{\nu_{\tau}}. \label{f1}
\end{align}

Supplementing 4-2-2 with a discrete left-right (LR) symmetry
\cite{pati,lr} (more precisely C-parity) \cite{c-parity} reduces the
number of independent gauge couplings in 4-2-2 from three to two.
This is because C-parity imposes the gauge
coupling unification condition ($g_L=g_R$) at $M_{\rm GUT}$. We will
assume that due to C-parity the SSB
mass terms, induced at $M_{\rm GUT}$ through gravity mediated
supersymmetry breaking \cite{Chamseddine:1982jx} are equal in magnitude for the  squarks
and sleptons of the three families. The tree level asymptotic MSSM
gaugino SSB masses, on the other hand, can be non-universal from the
following consideration. From C-parity, we can expect that the
gaugino masses at $M_{\rm GUT}$ associated with $SU(2)_L$ and
$SU(2)_R$ are the same ($M_2 \equiv M_2^R= M_2^L$). However, the
asymptotic $SU(4)_c$ and consequently $SU(3)_c$ gaugino SSB masses
can be different. With the hypercharge generator in 4-2-2 given by
$Y=\sqrt{2/5}~(B-L)+\sqrt{3/5}~I_{3R}$, where $B-L$ and $I_{3R}$ are
the diagonal generators of $SU(4)_c$ and $SU(2)_R$, we have the
following asymptotic relation between the three MSSM gaugino SSB
masses:
\begin{align}
M_1=\frac{3}{5} M_2 + \frac{2}{5} M_3. \label{gaugino-condition}
\end{align}

The supersymmetric 4-2-2 model with C-parity thus has two
independent parameters ($M_2$ and $M_3$) in the gaugino sector. In
order to implement Yukawa unification it turns out that the SSB Higgs
mass terms must be non-universal at $M_{\rm GUT}$. Namely,
$m_{H_u}^2<m_{H_d}^2$ at $M_{GUT}$, where $m_{H_u} (m_{H_d})$ is the
up (down) type SSB Higgs mass term. The fundamental parameters of
the 4-2-2 model that we consider are as follows:

\begin{align}
m_{0}, m_{H_u}, m_{H_d}, M_2, M_3, A_0, \tan\beta, {\rm sign}(\mu).
\label{params}
\end{align}
Here $m_0$ is the universal SSB mass for MSSM sfermions,
$A_0$ is the universal SSB trilinear scalar interaction (with the
corresponding Yukawa coupling factored out), $\tan\beta$ is the
ratio of the vacuum expectation values  (VEVs) of the two MSSM Higgs
doublets, and the magnitude of $\mu$, but not its sign, is determined by the
radiative electroweak breaking (REWSB) condition.
 In this paper we mainly focus on $\mu<0$ as well as
 $M_2<0$, as explained earlier. Although not required, we will
assume that the gauge coupling unification condition $g_3=g_1=g_2$
holds at $M_{\rm GUT}$ in 4-2-2. Such a scenario can arise,
for example, from a higher dimensional $SO(10)$
\cite{Hebecker:2001jb} or $SU(8)$ \cite{su8} model after suitable
compactification.

\section{Phenomenological constraints and scanning procedure\label{constraintsSection}}

We employ the ISAJET~7.80 package~\cite{ISAJET}  to perform random
scans over the parameter space listed in Eq.(\ref{params}). In this
package, the weak scale values of gauge and third generation Yukawa
couplings are evolved to $M_{\rm GUT}$ via the MSSM renormalization
group equations (RGEs) in the $\overline{DR}$ regularization scheme.
We do not strictly enforce the unification condition $g_3=g_1=g_2$ at $M_{\rm
GUT}$, since a few percent deviation from unification can be
assigned to unknown GUT-scale threshold
corrections~\cite{Hisano:1992jj}.
The deviation between $g_1=g_2$ and $g_3$ at $M_{GUT}$ is no
worse than $3.5\%$, and it is also possible to get
perfect gauge coupling unification. Perfect gauge coupling unification is
typically not possible with gaugino universality.
With nonuniversality in the gaugino sector, one can adjust
$M_3/M_2$ appropriately to get exact gauge coupling unification.
If neutrinos acquire mass via Type I seesaw, the impact of the neutrino Dirac
Yukawa coupling in the RGEs of the SSB terms, gauge couplings and the
third generation Yukawa
couplings is significant only for relatively large values ($\sim 2$ or so).
In the 4-2-2 model we expect the largest Dirac Yukawa coupling to be
comparable to the top Yukawa coupling ($\sim 0.6$ at $M_{\rm GUT}$).
Therefore, we do not include the Dirac neutrino Yukawa coupling
in the RGEs.

The various boundary conditions are imposed at
$M_{\rm GUT}$ and all the SSB
parameters, along with the gauge and Yukawa couplings, are evolved
back to the weak scale $M_{\rm Z}$.
In the evaluation of Yukawa couplings the SUSY threshold
corrections~\cite{Pierce:1996zz} are taken into account at the
common scale $M_{\rm SUSY}= \sqrt{m_{{\tilde t}_L}m_{{\tilde t}_R}}$. The entire
parameter set is iteratively run between $M_{\rm Z}$ and $M_{\rm
GUT}$ using the full 2-loop RGEs until a stable solution is
obtained. To better account for leading-log corrections, one-loop
step-beta functions are adopted for gauge and Yukawa couplings, and
the SSB parameters $m_i$ are extracted from RGEs at multiple scales
$m_i=m_i(m_i)$. The RGE-improved 1-loop effective potential is
minimized at an optimized scale $M_{\rm SUSY}$, which effectively
accounts for the leading 2-loop corrections. Full 1-loop radiative
corrections are incorporated for all sparticle masses.

The requirement of REWSB~\cite{Ibanez:1982fr} puts an important theoretical
constraint on the parameter space. Another important constraint
comes from limits on the cosmological abundance of stable charged
particles~\cite{PDG}. This excludes regions in the parameter space
where charged SUSY particles, such as ${\tilde \tau}_1$ or ${\tilde t}_1$, become
the lightest supersymmetric particle (LSP). We accept only those
solutions for which one of the neutralinos is the LSP and saturates
the WMAP (Wilkinson Microwave Anisotropy Probe) dark matter relic abundance bound.

We have performed random scans for the following parameter range:

\begin{align}0\leq  m_{0}, m_{H_u}, m_{H_d} \leq 20\, \rm{TeV} \nonumber \\
-2\rm{TeV} \leq M_2  \leq 0 \nonumber \\
0 \leq M_3  \leq 2\, \rm{TeV} \nonumber \\
45\leq \tan\beta \leq 55 \nonumber \\
-3\leq A_{0}/m_0 \leq 3\nonumber \\
\mu < 0,
 \label{parameterRange}
\end{align}
with $m_t = 173.1\, {\rm GeV}$ \cite{:2009ec}. The results are not
too sensitive to one or two sigma variation in the value of $m_t$.
We use $m_b(m_Z)=2.83$ GeV which is hard-coded into ISAJET. This
choice of parameters was influenced by our previous experience with
the 4-2-2 where we set $\mu,M_2>0$.

In scanning the parameter space, we employ the Metropolis-Hastings
algorithm as described in \cite{Belanger:2009ti}. All of the
collected data points satisfy
the requirement of REWSB,
with the neutralino in each case being the LSP. Furthermore,
all of these points satisfy the constraint $\Omega_{\rm CDM}h^2 \le 10$.
This is done so as to collect more points with a WMAP compatible value of cold dark
matter (CDM) relic abundance. For the Metropolis-Hastings algorithm, we only use
the value of $\Omega_{\rm CDM}h^2$ to bias our search. Our purpose in using the
Metropolis-Hastings algorithm is to be able to search around regions of
acceptable $\Omega_{\rm CDM}h^2$ more fully. After collecting the data, we impose
the mass bounds on all the particles~\cite{Amsler:2008zzb} and use the
IsaTools package~\cite{Baer:2002fv}
to implement the following phenomenological constraints:
\begin{table}[h]\centering
\begin{tabular}{rlc}
$m_h~{\rm (lightest~Higgs~mass)} $&$ \geq\, 114.4~{\rm GeV}$                    &  \cite{Schael:2006cr} \\
$BR(B_s \rightarrow \mu^+ \mu^-) $&$ <\, 5.8 \times 10^{-8}$                     &   \cite{:2007kv}      \\
$2.85 \times 10^{-4} \leq BR(b \rightarrow s \gamma) $&$ \leq\, 4.24 \times 10^{-4} \; (2\sigma)$ &   \cite{Barberio:2008fa}  \\
$0.15 \leq \frac{BR(B_u\rightarrow \tau \nu_{\tau})_{\rm MSSM}}{BR(B_u\rightarrow \tau \nu_{\tau})_{\rm SM}}$&$ \leq\, 2.41 \; (3\sigma)$ &   \cite{Barberio:2008fa}  \\
$\Omega_{\rm CDM}h^2 $&$ =\, 0.111^{+0.028}_{-0.037} \;(5\sigma)$               &  \cite{Komatsu:2008hk}    \\
$3.4 \times 10^{-10}\leq \Delta (g-2)_{\mu}/2 $&$ \leq\, 55.6 \times 10^{-10}~ \; (3\sigma)$ &  \cite{Bennett:2006fi}
\end{tabular}

\end{table}

We apply the experimental constraints successively on the data that
we acquire from ISAJET.

\section{Sign of $\mu$ and Yukawa unification\label{muneg}}

\begin{figure}[!hb]
\centering \subfiguretopcaptrue

\includegraphics[width=8cm]{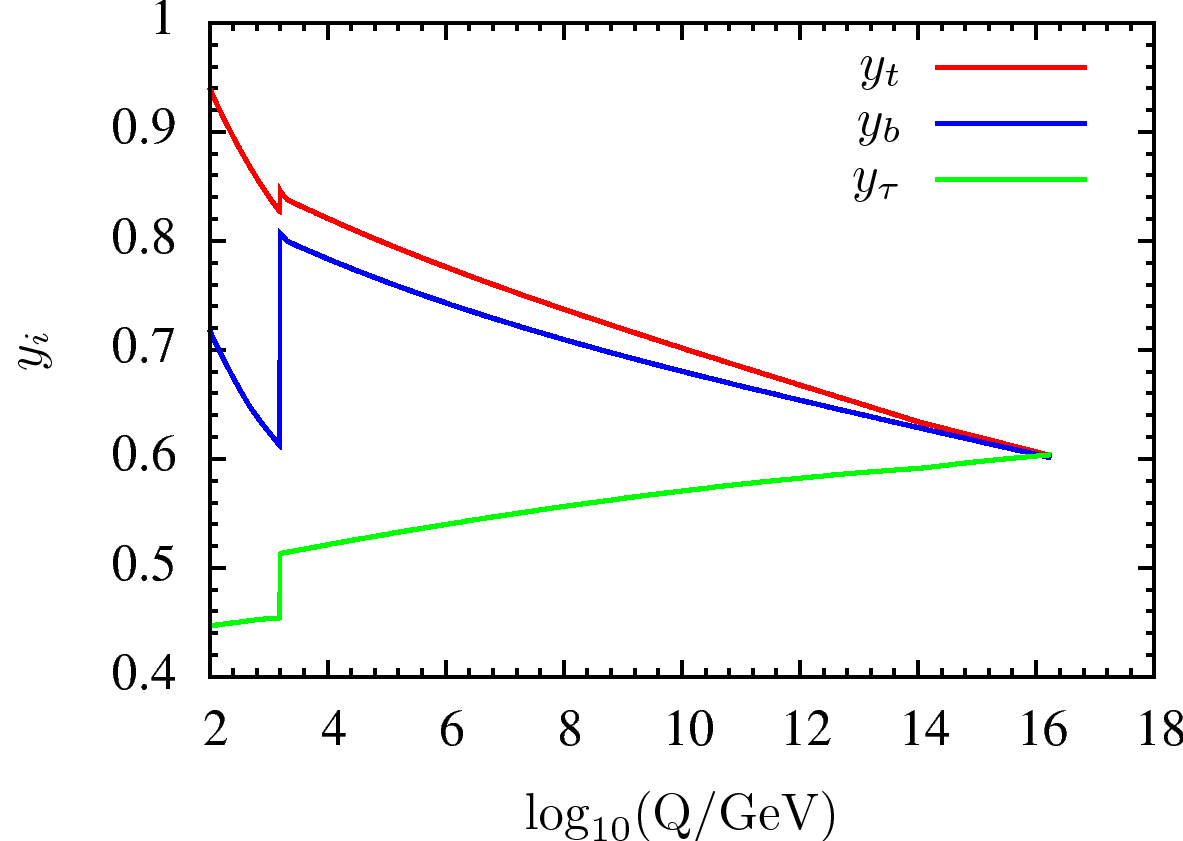}

\caption{Evolution of top (red), bottom (blue) and tau (green)
Yukawa couplings for  $\mu<0$.} \label{compareMunegMupos}
\end{figure}

To first appreciate the impact of the sign of $\mu$ on Yukawa
coupling unification and to see why $\mu<0$ is preferred over $\mu>0$,
in Fig.~\ref{compareMunegMupos} we show the evolution of the top, bottom and tau
Yukawa couplings for a representative Yukawa coupling unification
solution. We observe that Yukawa unification requires relatively large
threshold corrections to $y_b$. To quantify the magnitude of the threshold
corrections, we scan the parameter space given in Eq.(\ref{parameterRange}) and
calculate the finite and logarithmic corrections to $\delta y_i$, where the index $i$ refers
to top, bottom and tau Yukawa couplings.

\begin{figure}[t!]
\centering \subfiguretopcaptrue
 \subfigure[\hspace {1mm}  $\mu>0$]{
\includegraphics[width=8cm]{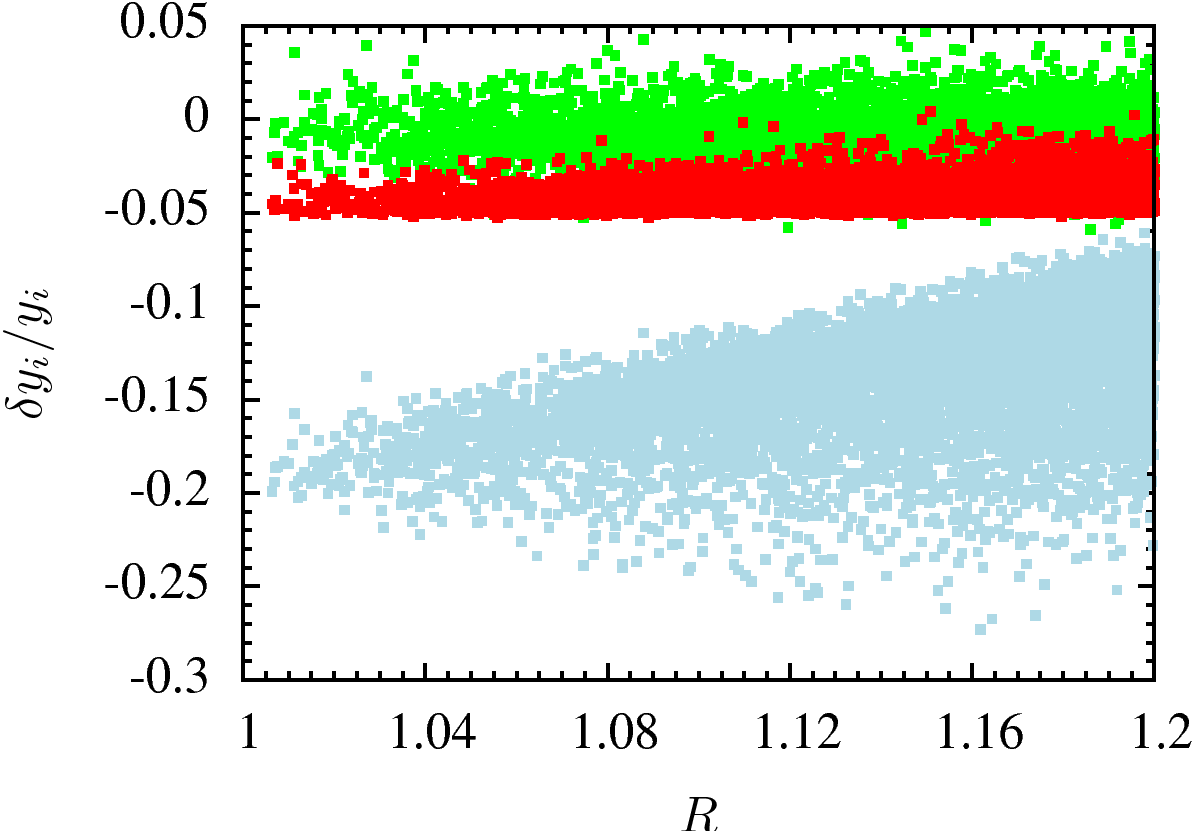}
\label{deltayiR-a} } \subfigure[\hspace {1mm}  $\mu<0$]{
\includegraphics[width=8cm]{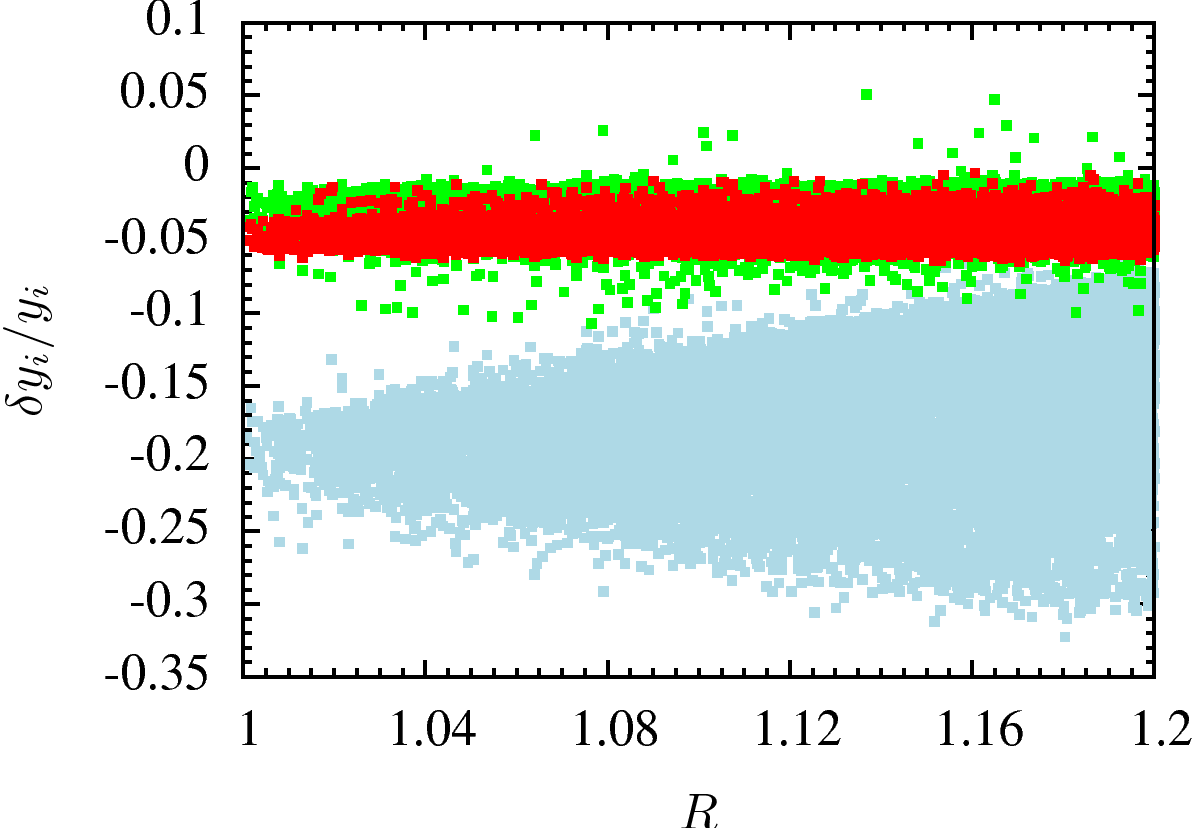}
\label{deltayiR-b} }

\caption{$\delta y_b/y_b$ (blue), $\delta y_{\tau}/y_{\tau}$ (green)
and $\delta y_t/y_t$ (red) versus $R$ (measure of Yukawa coupling
unification).} \label{deltayiR}
\end{figure}

Following \cite{Baer:2008jn}, we define the quantity $R$ as,

\begin{align}
R=\frac{ \rm max(y_t,y_b,y_{\tau})} { {\rm min} (y_t,y_b,y_{\tau})}
\end{align}
Thus, $R$ is a useful indicator for Yukawa unification with $R\leq 1.1$,
for instance, corresponding to Yukawa unification within 10\%.

Fig.~\ref{deltayiR} shows a plot of  $\delta y_i/y_i$ versus $R$
for both $\mu>0$ and
$\mu<0$. Points in blue, green and red represent, respectively,
$\delta y_b/y_b$, $\delta y_{\tau}/y_{\tau}$ and $\delta y_t/y_t$.
Note that we choose the sign of  $\delta y_i$ from the
perspective of evolving $y_i$ from $M_{\rm GUT}$ to $M_{\rm Z}$.
Fig.~\ref{deltayiR} confirms the trend seen in
Fig.~\ref{compareMunegMupos} that $y_t$ and $y_{\tau}$ receive small
threshold corrections compared to $y_b$. Therefore, it is reasonable
to focus on the threshold corrections to $y_b$ while studying Yukawa
unification.

\begin{figure}[!b]
\centering \subfiguretopcaptrue

\subfigure[\hspace {1mm}  $\mu>0$]{
\includegraphics[width=8cm]{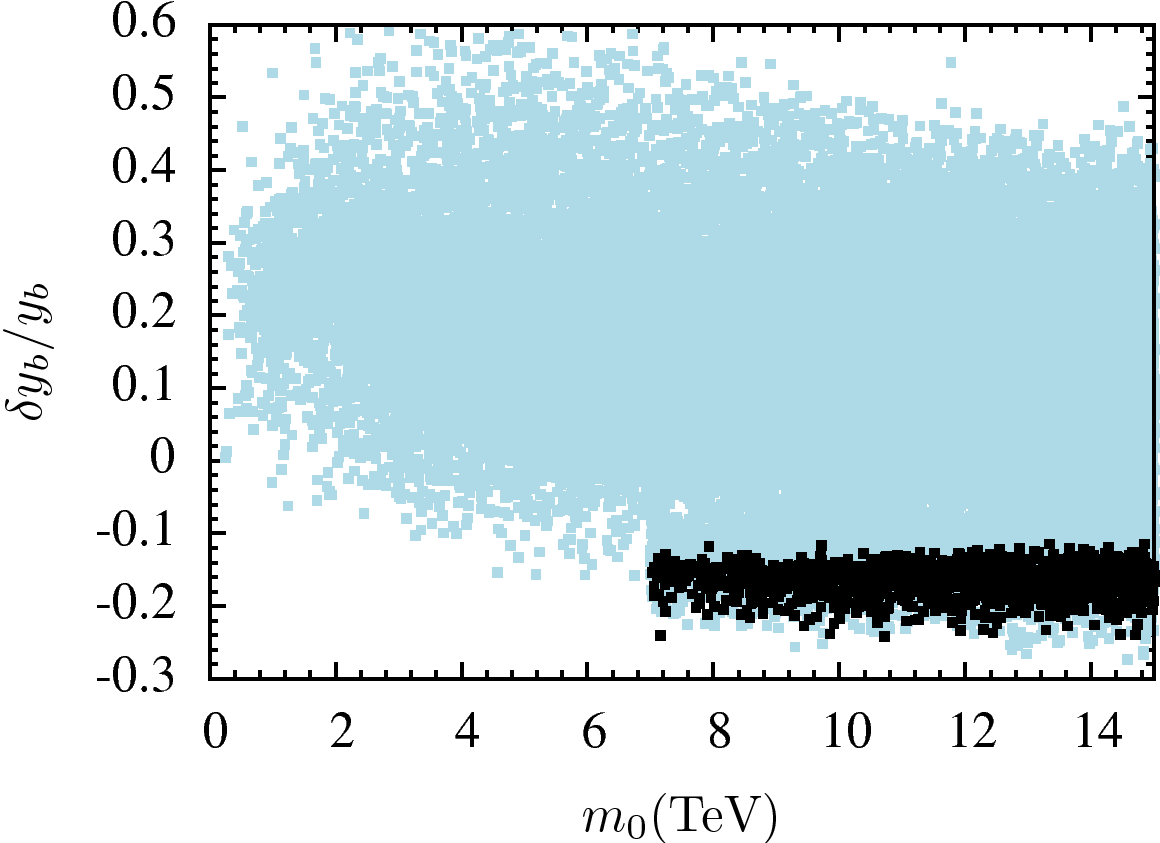}
\label{deltaybm0-a} } \subfigure[\hspace {1mm}  $\mu<0$]{
\includegraphics[width=8cm]{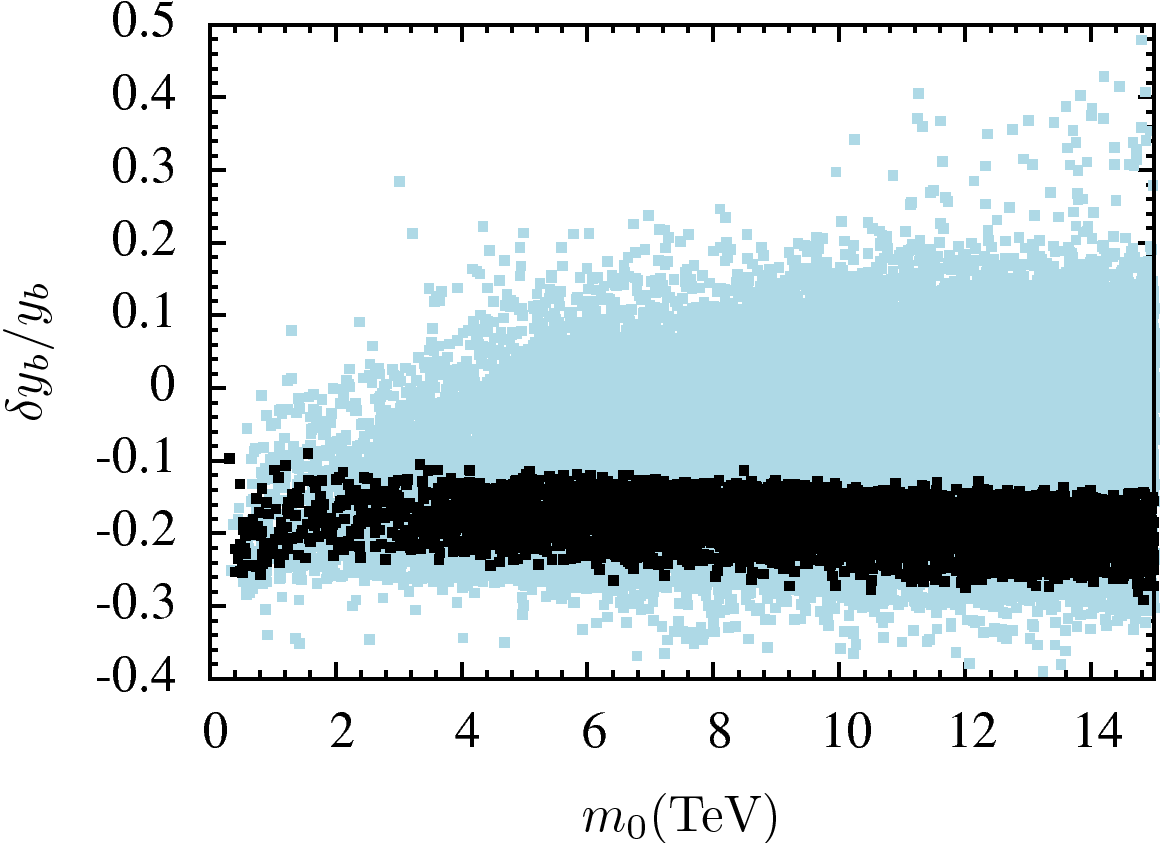}
\label{deltaybm0-b} }

\caption{$\delta y_b/y_b$ as a function of $m_0$. Points in black
correspond to 10\% or better Yukawa unification ($R<=1.1$).}
\label{deltaybm0}
\end{figure}

The scale at which Yukawa coupling unification is to occur is set by
gauge coupling unification and is $M_{\rm GUT}$. Let us first
consider the case of $y_t ( M_{\rm GUT})\approx y_{\tau}(M_{\rm
GUT})$. Because the threshold corrections to $y_t$ are very small,
it is convenient to think of $y_{\tau}$. The SUSY correction to the
tau lepton mass $\delta m_{\tau}$ is given by $\delta
m_{\tau}=v\cos\beta \delta y_{\tau}$. In order to get the correct
$\tau$ mass ($m_{\tau}$), one has to get an appropriate $\delta
y_{\tau}$. Because of the range of values of $\cos\beta$ for large
$\tan\beta$, there is freedom to choose the value of $\delta
y_{\tau}$. It may be possible to trade this freedom in favor of
top-tau Yukawa unification $y_t (M_{\rm GUT})\approx y_{\tau}(M_{\rm
GUT})$. One then needs the correct SUSY contribution to $\delta y_b$
in order to achieve Yukawa coupling unification $y_t (M_{\rm
GUT})\approx y_{b}(M_{\rm GUT}) \approx y_{\tau}(M_{\rm GUT})$.

In order to understand why sign of $\mu$ is very crucial for Yukawa
unification condition lets first analyze the analytical expression of
threshold corrections for the bottom Yukawa coupling.
 The dominant contribution to $\delta y_b$ comes from
the gluino and chargino loops, and in our sign convention, is given by
~\cite{Pierce:1996zz}

\begin{align}
\delta y_b^{\rm finite}=\frac{g_3^2}{12\pi^2}\frac{\mu m_{\tilde g}
\tan\beta}{m_{\tilde b}^2}+
                         \frac{y_t^2}{32\pi^2}\frac{\mu A_t \tan\beta}{m_{\tilde t}^2},
\end{align}
where $g_3$ is the strong gauge coupling constant, $m_{\tilde g}$ is
the gluino mass, $m_{\tilde b}$ is the sbottom mass, $m_{\tilde t}$
is the stop mass, and $A_t$ is the top trilinear coupling.
One can see from Fig.~\ref{deltayiR}
that in order to achieve Yukawa coupling unification $R\sim 1$, the
threshold corrections to $y_b$ have to be negative (in our sign
convention for $\delta y_i$) and in a somewhat narrow interval
($-0.5 \lesssim \delta y_b/y_b \lesssim -1.5$) considering the full
range of possible values of $\delta y_b$. The logarithmic
corrections to $y_b$ are in fact positive.
This leaves the finite corrections to provide for the correct
$\delta y_b$ to compensate for the `wrong' sign of the logarithmic
corrections. If $\mu>0$, the gluino contribution is positive, and so the
contribution from the chargino loop must cancel the contribution
from the gluino loop and the logarithmic correction, as well as provide
the correct (negative) contribution to $\delta y_b$. This can be
achieved only for a large $m_0$, as for large $m_0$ and for large
$A_t$, the gluino contribution scales as $M_{1/2}/m_0^2$ while the
chargino contribution scales as $A_t/m_0^2$. It also should be noted
that the numerical factor for the gluino contribution is larger than
than the corresponding factor for chargino contribution. Therefore, a sufficiently large
value of $A_t$ and $m_0$ is needed. This large required value of
$A_t$ is the reason behind the requirement of $A_0/m_0\sim -2.6$ for
$\mu>0$.

The scenario with $\mu<0$ is interesting because the gluino
contribution to $\delta y_b$ has the correct sign to
obtain the required b-quark mass. Thus, we should
expect that with $\mu<0$, we can realize Yukawa unification for
a wider range $A_0$ values. With the threshold contribution to $y_{\tau}$
proportional to $M_2 \tan\beta$, and with $M_2$ as a
free parameter in the 4-2-2 model, we also should expect
Yukawa unification to occur over a broader range of
$\tan\beta$ values.

To proceed further, in Fig. \ref{deltaybm0} we present
$\delta y_b/y_b$ versus $m_0$ for $\mu>0$ and $\mu<0$, by performing
a scan over the parameter space given in Eq.(\ref{parameterRange}).
Here $\delta y_b$ includes full one loop finite and logarithmic
corrections, and the black points correspond to 10\% or better Yukawa
unification ($R\leq 1.1$). It is clear from Fig. \ref{deltaybm0}
that Yukawa coupling unification with a relatively light
 $m_0\sim400$ GeV can be realized for $\mu<0$. The  $\mu>0$ scenario typically
requires a very large $m_0\gtrsim 8$  TeV.

\begin{figure}[t]
\centering
\subfiguretopcaptrue

\subfigure{
\includegraphics[width=8cm]{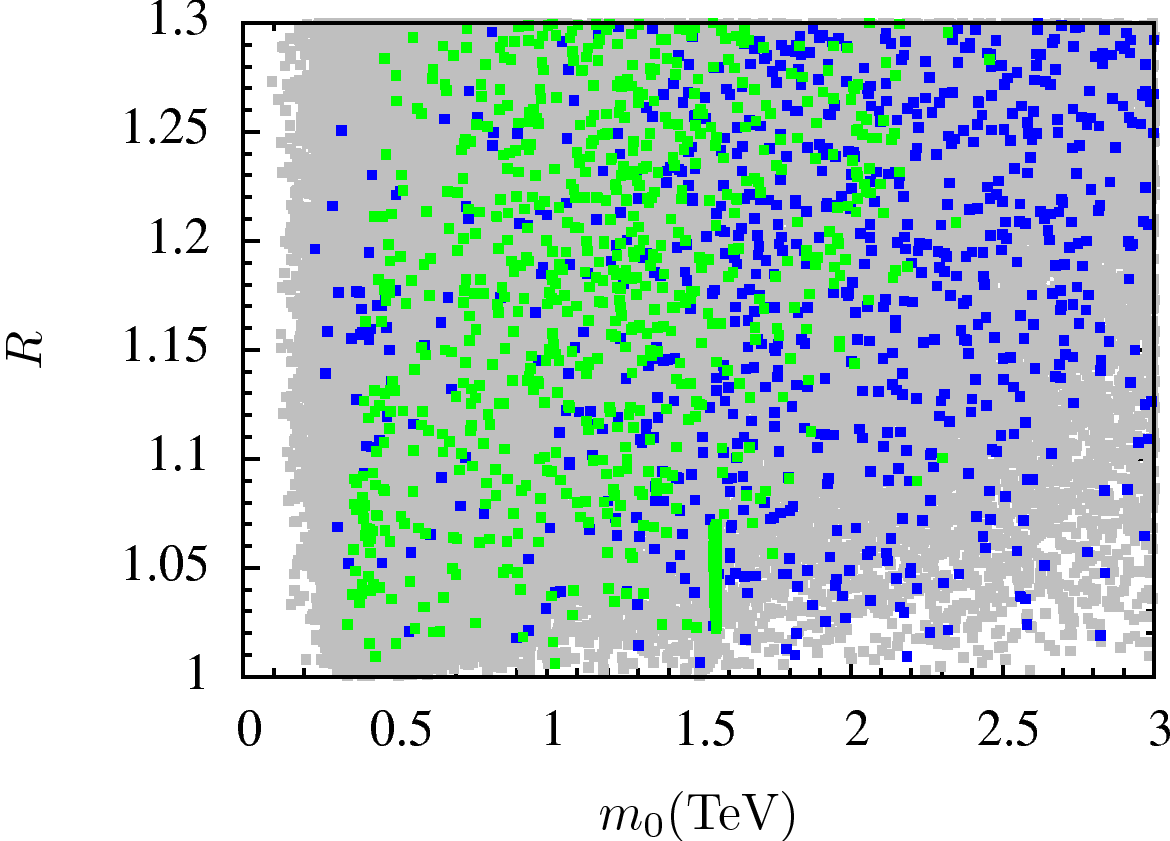}
}
\subfigure{
\includegraphics[width=8cm]{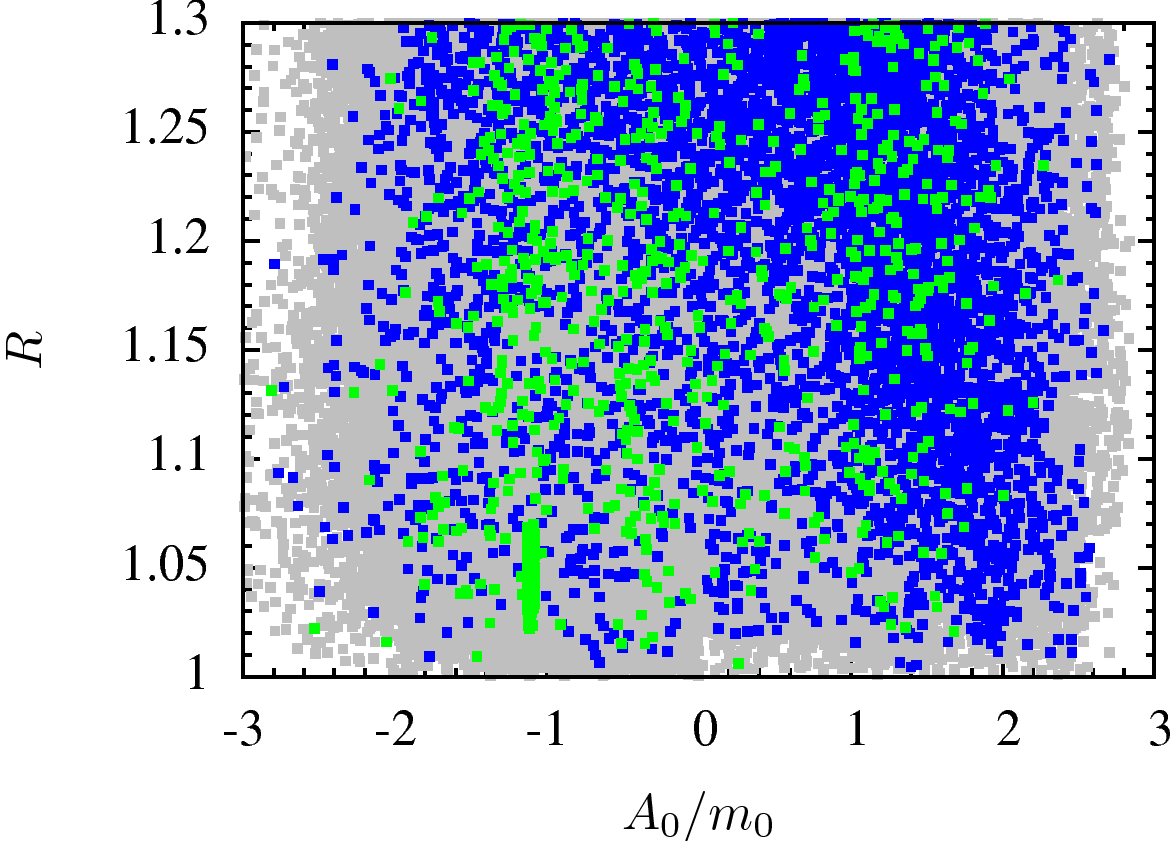}
}
\subfigure{
\includegraphics[width=8cm]{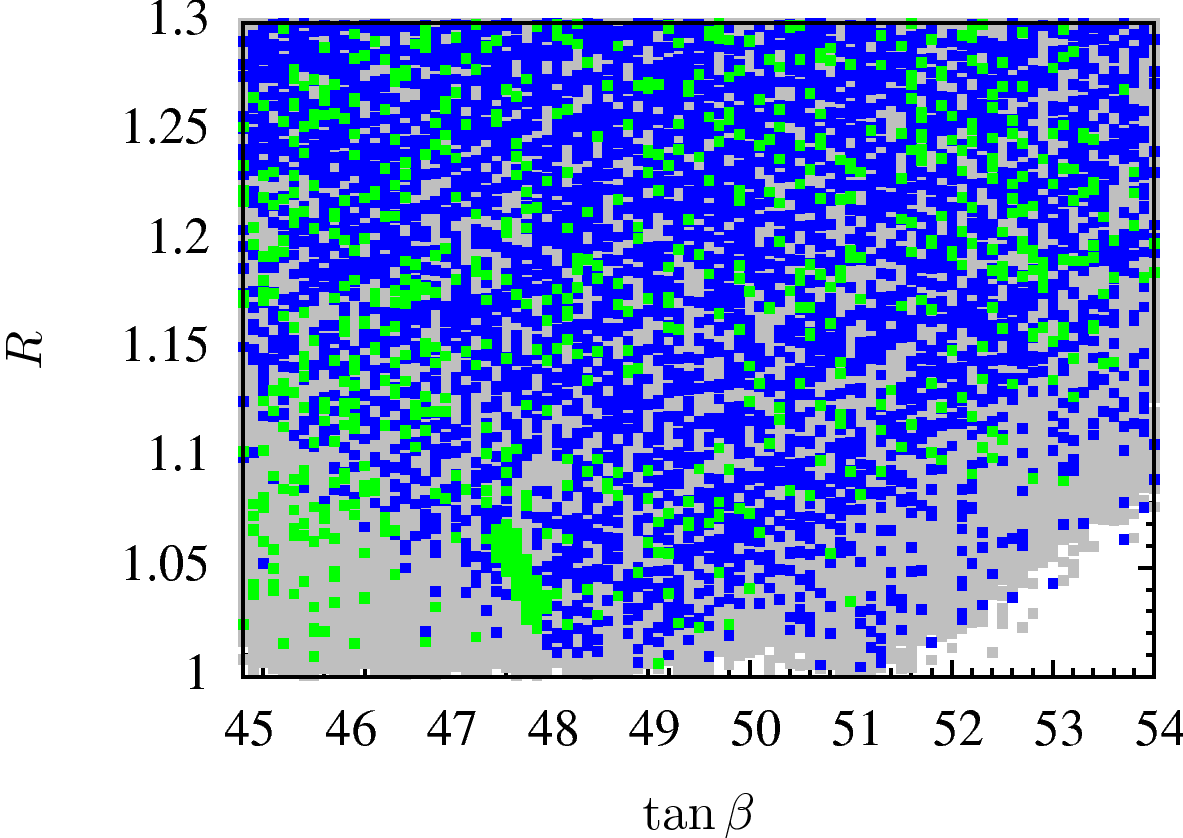}
}
\subfigure{
\includegraphics[width=8cm]{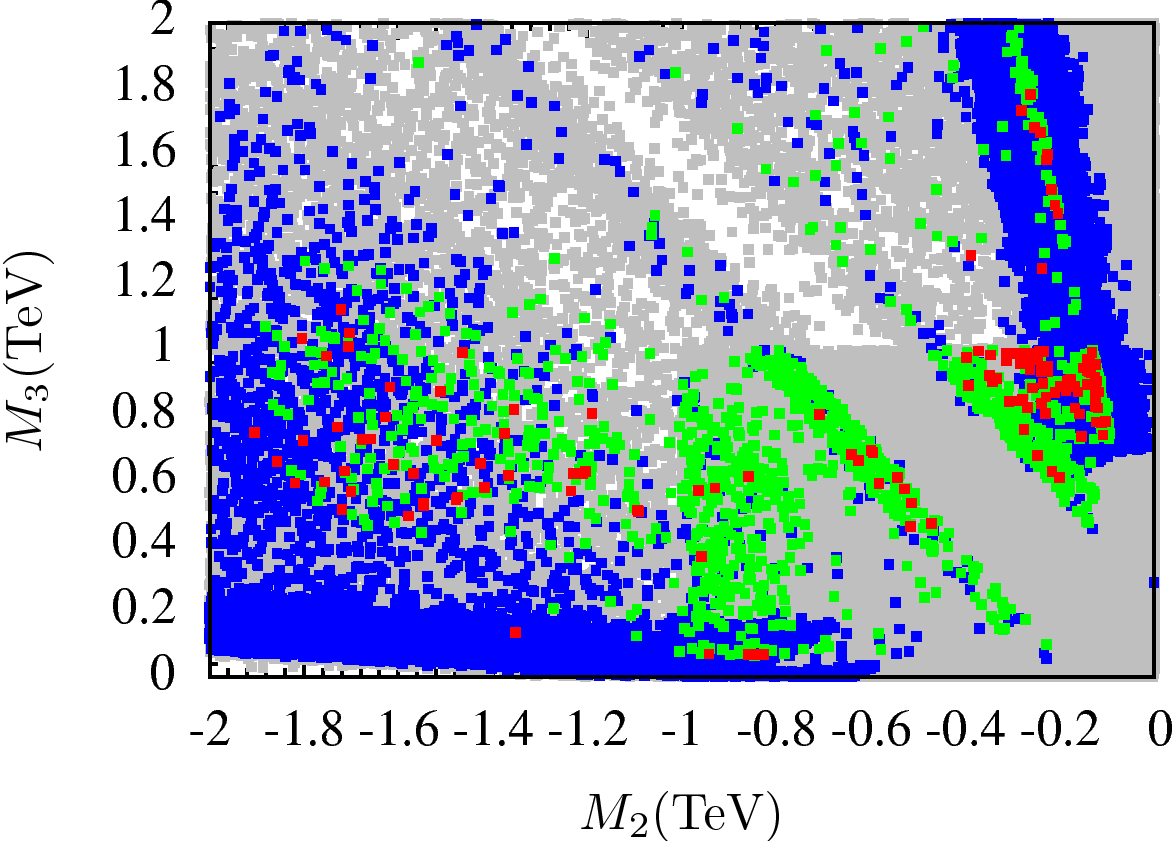}
}

\caption{
Plots in the $R$ - $m_0$, $R$ - $\tan\beta$, $R$ -
$A_0/m_0$ and $M_3$ - $M_2$ planes. Gray points are consistent with
REWSB and $\tilde{\chi}^0_{1}$  LSP. Blue points satisfy the WMAP
bounds on $\tilde{\chi}^0_1$ dark matter abundance particle mass
bounds, constraints from $BR(B_s\rightarrow \mu^+ \mu^-)$,
$BR(B_u\rightarrow \tau \nu_{\tau})$ and
$BR(b\rightarrow s \gamma)$. Green points belong to the subset of
blue points that satisfies all constraints including $(g-2)_\mu$. In
the $M_3$ - $M_2$ plane, points in red represent the subset of green
points that satisfies Yukawa coupling unification to within 10\%.
\label{fund}
}
\end{figure}

\section{Yukawa unification and sparticle spectroscopy \label{results}}

We now present the results of the scan over the parameter space
listed in Eq.(\ref{parameterRange}). In Fig.~\ref{fund} we show the
results in the $R$ - $m_0$, $R$ - $\tan\beta$, $R$ - $A_0/m_0$ and
$M_3$ - $M_2$ planes. The gray points are consistent with REWSB and
$\tilde{\chi}^0_{1}$  LSP. The blue points satisfy the WMAP bounds on
$\tilde{\chi}^0_1$ dark matter abundance, sparticle mass bounds,
constraints from $BR(B_s\rightarrow \mu^+ \mu^-)$ and
$BR(b\rightarrow s \gamma)$. The green points belong to the subset of
blue points that satisfy all constraints including $(g-2)_\mu$. In
the $M_3$ - $M_2$ plane, points in red represent the subset of green
points that satisfies Yukawa coupling unification to within 10\%.

In the $R$ - $m_0$ plane of Fig.~\ref{fund} we see that with both
$\mu<0$ and $M_2<0$, we can realize
Yukawa unification consistent with all constraints mentioned in
Section~\ref{constraintsSection} including the one from $
(g-2)_{\mu}$. This is possible, as previously noted, because we can now implement Yukawa
unification for relatively small $m_0$($\sim 400\, {\rm GeV}$) values because $\mu<0$,
and, in turn, $(g-2)_{\mu}$ obtains the desired SUSY contribution
which is proportional to $\mu M_2$. This is more than an
order of magnitude improvement on the $m_0$ value required for
Yukawa unification with  $\mu>0$. We also see from the $R$ - $A_0/m_0$ plane that, as
explained earlier, Yukawa unification for an essentially
arbitrary value of $A_0$ is obtained for $\mu<0$. Our observation about relaxing
the possible range of $\tan\beta$ that accommodates Yukawa unified
models is explicitly shown in the $R$ - $\tan\beta$ plane. This
also suggests that we are perhaps somewhat conservative in limiting the range of
$\tan\beta$. While it is beyond the scope of this paper, it would be nice to systematically
search for a lower bound on $\tan\beta$ that is consistent with third family Yukawa unification.

\begin{figure}[!ht]
\centering
\subfiguretopcaptrue

\subfigure{
\includegraphics[width=8cm]{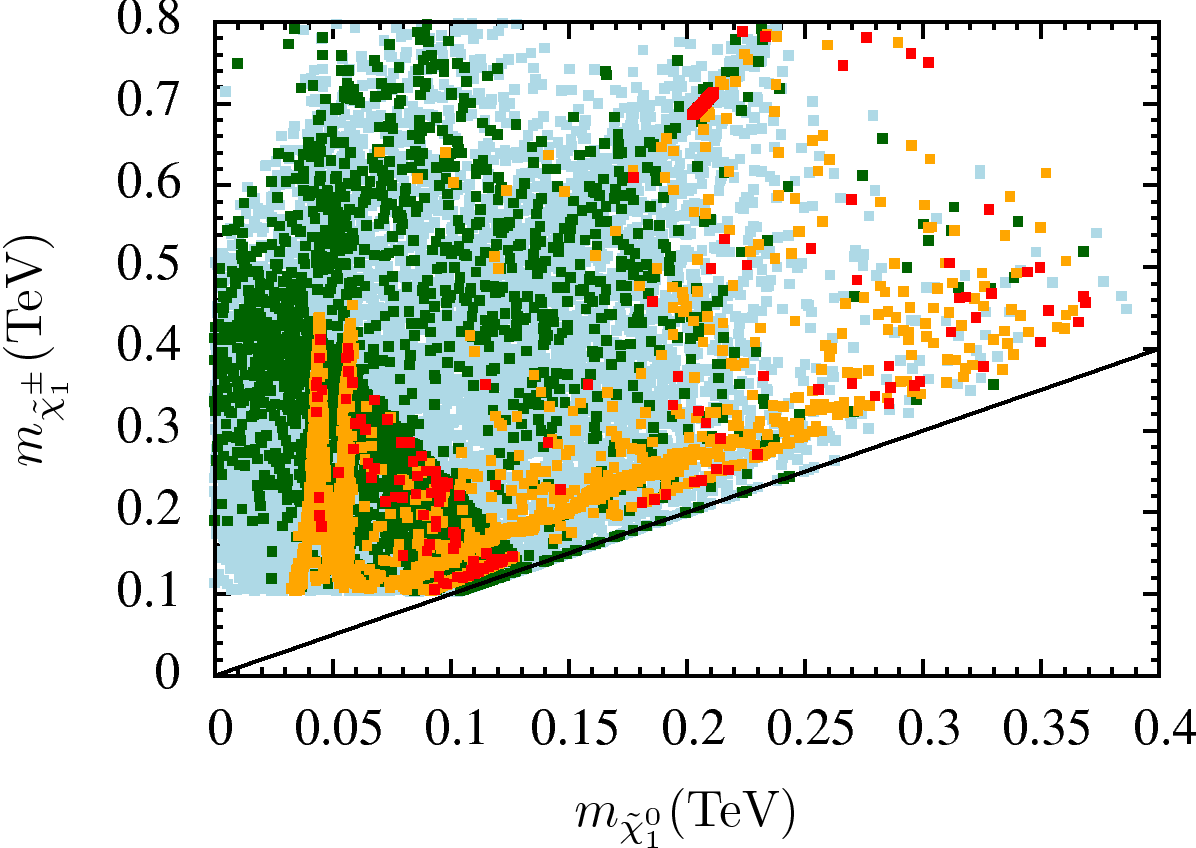}
}
\subfigure{
\includegraphics[width=8cm]{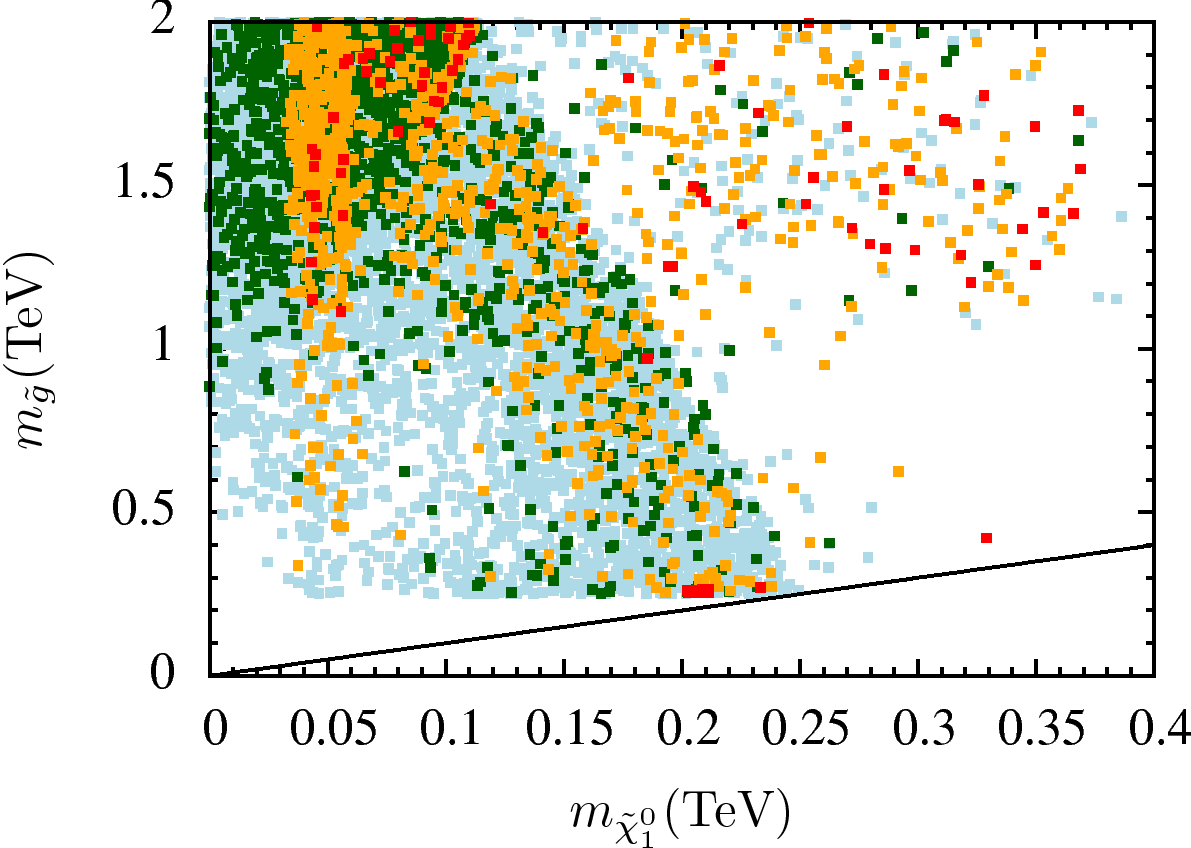}
}
\subfigure{
\includegraphics[width=8cm]{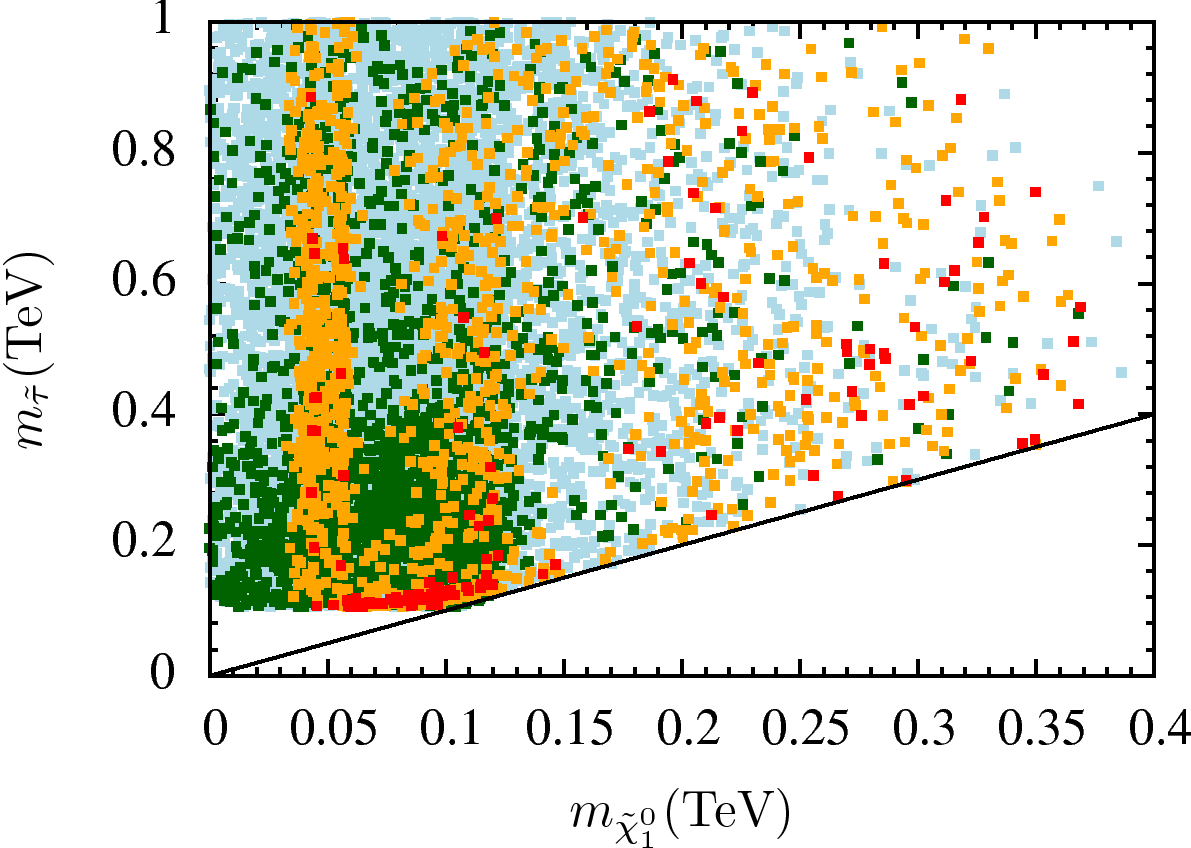}
}
\subfigure{
\includegraphics[width=8cm]{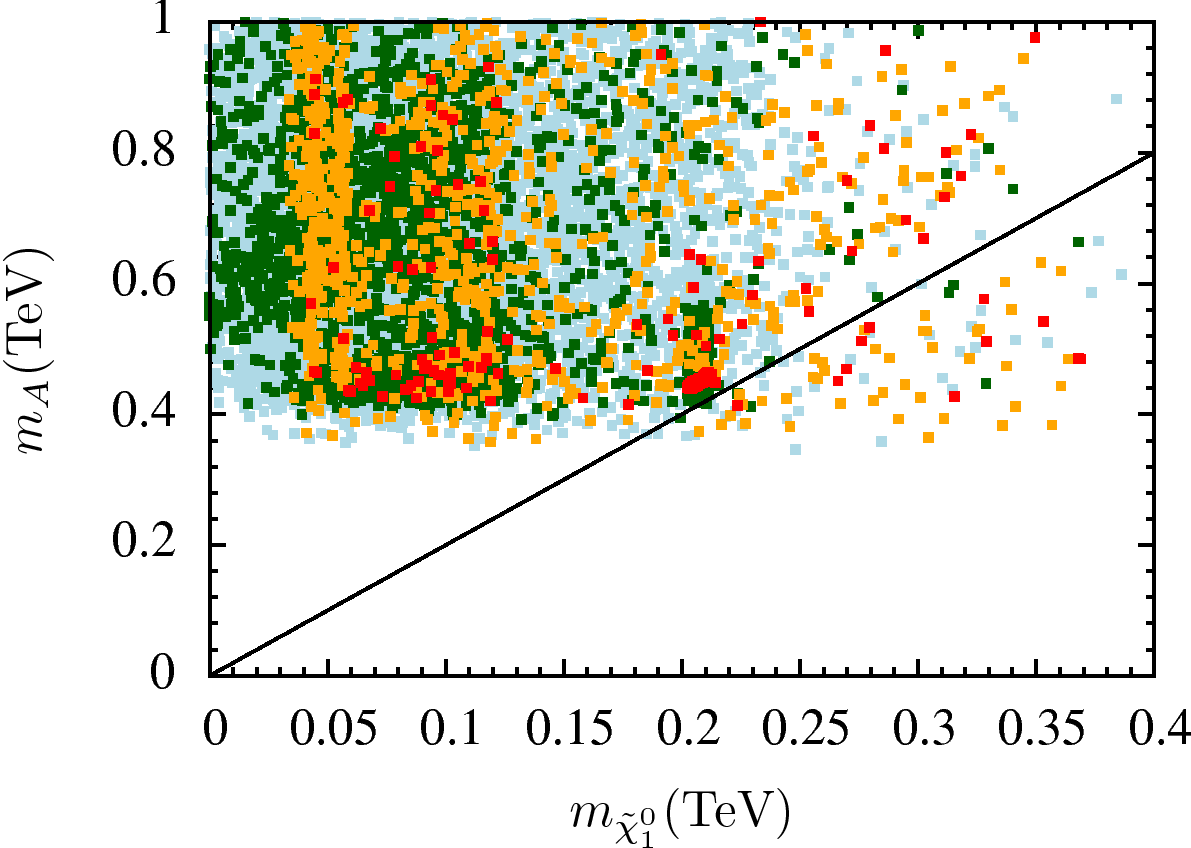}
}

\caption{
Plots in the $m_{\tilde{\chi}_1^{\pm}}$ - $m_{\tilde{
\chi}_1^0}$, $m_{\tilde {g}}$ - $m_{\tilde{ \chi}_1^0}$, $m_{\tilde
{\tau}}$ - $m_{\tilde{ \chi}_1^0}$ and $m_A$ - $m_{\tilde{
\chi}_1^0}$ planes. All points satisfy the requirements of REWSB,
$\tilde{\chi}^0_{1}$ LSP,  particle mass bounds and constraints from
$BR(B_s\rightarrow \mu^+ \mu^-)$,
$BR(B_u\rightarrow \tau \nu_{\tau})$ and $BR(b\rightarrow s \gamma)$.
Light blue points further satisfy the constraint on $(g-2)_\mu$.
Green points form a subset of light blue points that satisfies
Yukawa unification to within 10\%. Orange points satisfy all the
constraints mentioned in Section~\ref{constraintsSection} while
red points form a subset of orange points that have $R\leq1.1$.
\label{spectra}
}
\end{figure}

The $M_3$ - $M_2$ plane of Fig.~\ref{fund} has some interesting features. The
large gray regions appear because the relic density of neutralinos
is too high. This may seem peculiar given that we have
non-universal Higgs boundary conditions. However, it is
readily explained by the fact that $M_2<0$.
The bulk of the gray region has a neutralino that is too light
($m_{\tilde{ \chi}_1^0}\lesssim 30\, {\rm GeV}$). While it appears
possible to have a relic density of a relatively light neutralino
 ($\lesssim 30\, {\rm GeV}$) consistent with WMAP, $BR(B_s\rightarrow \mu^+\mu^-)$ and
$BR(b\rightarrow s \gamma)$, so that some of this gray region may turn
blue, there will always be regions where some of these constraints are not satisfied.

In Fig.~\ref{spectra} we show the relic density channels consistent
with Yukawa unification in the $m_{\tilde{\chi}_1^{\pm}}$ -
$m_{\tilde{ \chi}_1^0}$, $m_{\tilde {g}}$ - $m_{\tilde{ \chi}_1^0}$,
$m_{\tilde {\tau}}$ - $m_{\tilde{ \chi}_1^0}$ and $m_A$ -
$m_{\tilde{ \chi}_1^0}$ planes. All of the points shown in this figure
satisfy the requirements of REWSB, $\tilde{\chi}^0_{1}$ LSP,
particle mass bounds and constraints from $BR(B_s\rightarrow \mu^+
\mu^-)$ and $BR(b\rightarrow s \gamma)$. The light blue points
satisfy, in addition the constraint from $(g-2)_\mu$. The green points form a subset of
light blue points that satisfies Yukawa unification to within 10\%.
The orange points satisfy all the constraints mentioned in
Section~\ref{constraintsSection}, while the red points form a subset of
orange points that have $R\leq1.1$. This choice of color coding is
influenced from displaying the sparticle spectrum with and without
neutralino dark matter, while still focussing on all the other
experimental constraints. The idea is to show the myriad of
solutions that implement Yukawa unification and are consistent with
all known experimental bounds except for the bound on relic dark matter
density from WMAP. The appearance of a variety of Yukawa unified solutions
with a very rich sparticle spectrum is a characteristic feature of $\mu<0$.

We can see in Fig.~\ref{spectra} that a variety of coannihilation
and annihilation scenarios are compatible with Yukawa unification
and neutralino dark matter. Included in the $m_A$ - $m_{\tilde{
\chi}_1^0}$ plane is the line $m_A$ = $2 m_{\tilde{ \chi}_1^0}$
which indicates that the $A$ funnel region is compatible
with Yukawa unification. In the remaining planes in
Fig.~\ref{spectra}, we draw the unit slope line which indicates the
presence of gluino, stau and wino coannihilation scenarios.
From the $m_{\tilde{\chi}_1^{\pm}}$ - $m_{\tilde{
\chi}_1^0}$ plane, it is easy to see the light Higgs ($h$) and
$Z$ resonance channels. Our results are focussed on relatively light
neutralinos ($m_{\tilde{\chi}_1^0}\lesssim225\, {\rm GeV}$). We expect, based on the
discussion presented, that other coannihilation channels like the
stop coannihilation scenario are also consistent with Yukawa
unification but we did not find them because of lack of statistics.

\section{Yukawa unification and dark matter detection   \label{dark}}

\begin{figure}[t!]
\centering
\subfiguretopcaptrue

\subfigure{
\includegraphics[width=8cm]{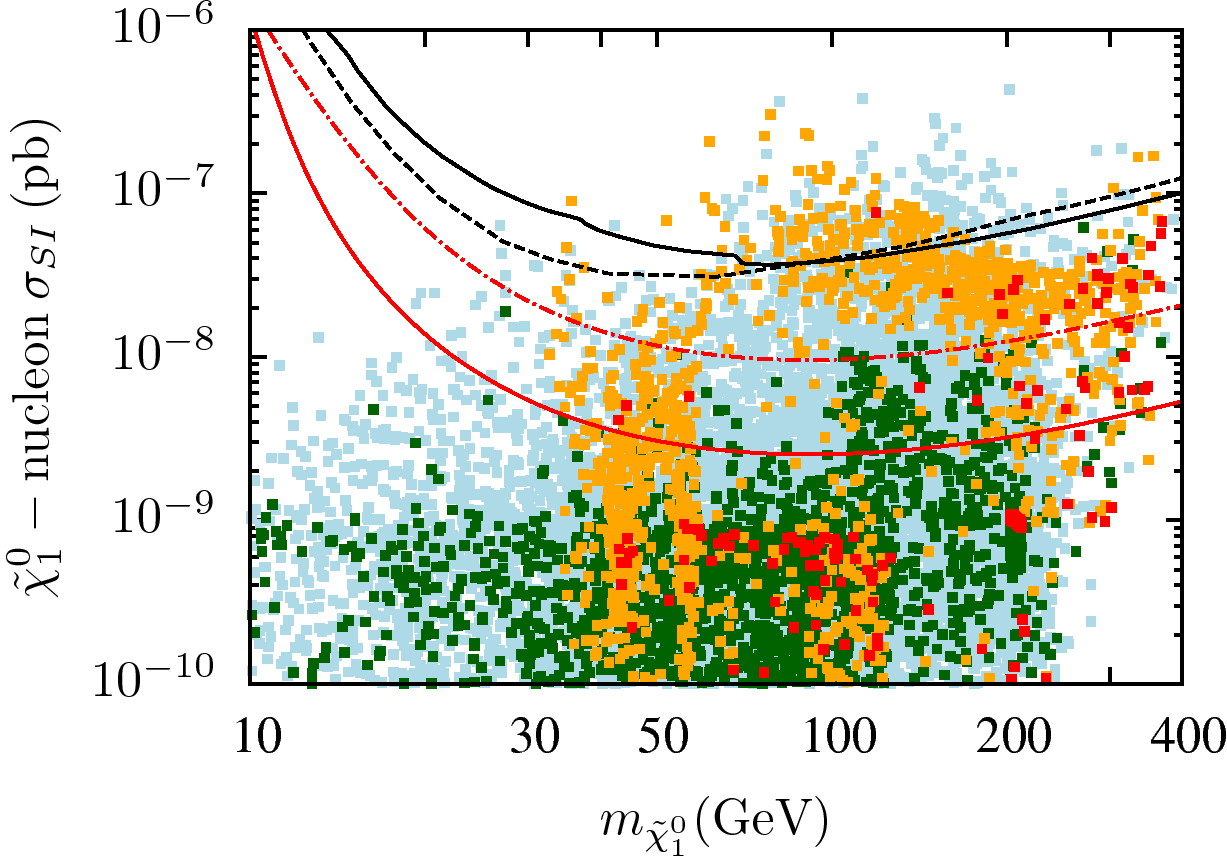}
}
\subfigure{
\includegraphics[width=8cm]{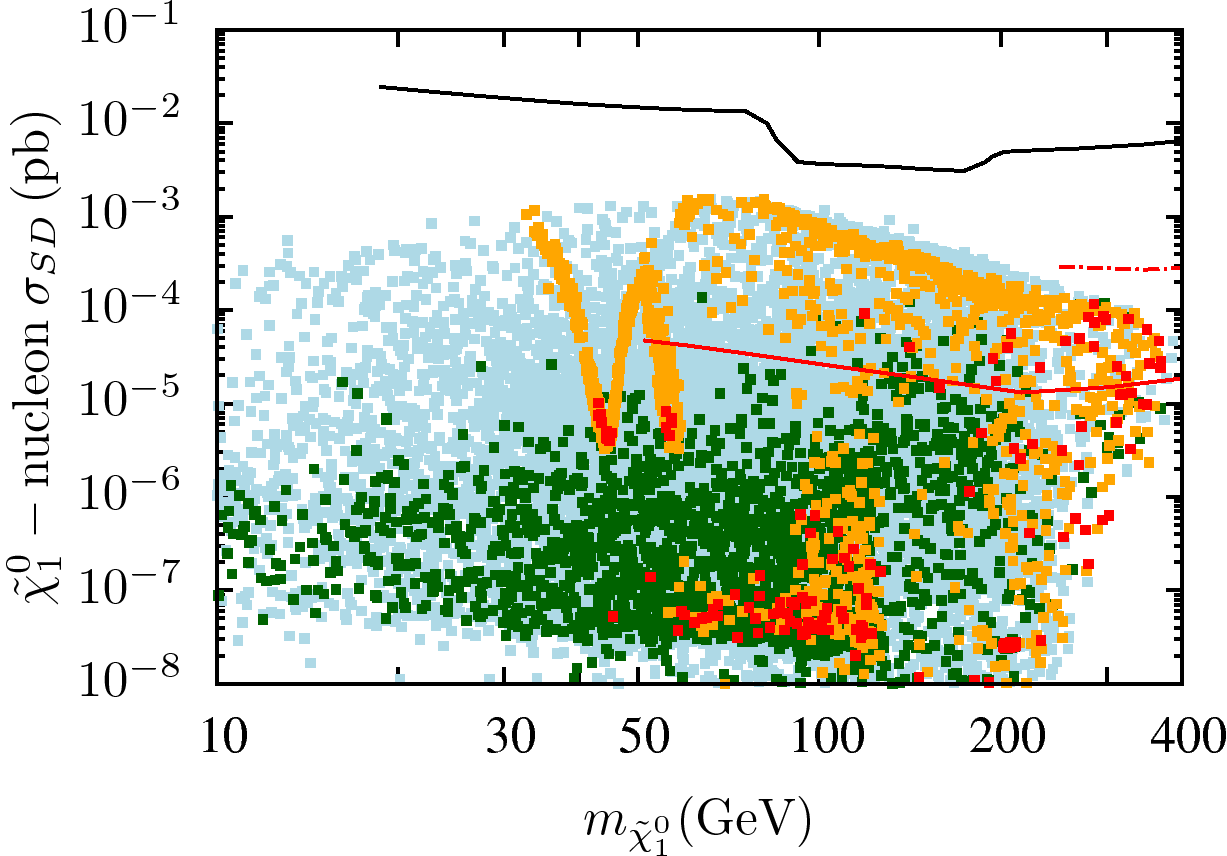}
}

\caption{Plots in the $\sigma_{\rm SI}$ - $m_{\tilde{\chi}_1^{0}}$
and $\sigma_{\rm SD}$- $m_{\tilde{\chi}_1^{0}}$ planes. Color coding
 is the same as in Fig.~\ref{spectra}. In the $\sigma_{\rm SI}$ -
$m_{\tilde{\chi}_1^{0}}$ plane we show the current bounds (black
lines) and future reaches (red lines) of the CDMS (solid lines) and
Xenon  (dotted lines) experiments. In the $\sigma_{\rm SD}$ -
$m_{\tilde{\chi}_1^{0}}$ plane we show the current bounds from Super
K (black line) and IceCube (dotted red line) and future reach of
IceCuce DeepCore (red solid line). \label{SIneuSDneu} }
\end{figure}

In light of the recent results by the CDMS-II \cite{Ahmed:2009zw}
and Xenon100 \cite{Aprile:2010um}  experiments, it is important to
see if Yukawa unification, within the framework presented in this
paper, is testable from the perspective of direct and indirect
detection experiments. The question of interest is whether $\mu \sim
M_1$ is consistent with Yukawa unification, as this is the
requirement to get a bino-higgsino admixture for the lightest
neutralino which, in turn, enhances both the spin dependent and spin
independent neutralino-nucleon scattering cross sections.
In Fig.~\ref{SIneuSDneu} we show the spin independent and spin
dependent cross sections as a function
of the neutralino mass. In the case of spin independent cross
section, we also show the current bounds and expected reach of the
CDMS and Xenon experiments. The color coding is the same as in
Fig.~\ref{spectra}. A small region of the parameter space consistent
with Yukawa unification and the experimental constraints discussed
in Section~\ref{constraintsSection} (red points in the
figure) is excluded as we can see, by the current CDMS and
XENON bounds. This shows that the ongoing and planned direct
detection experiments will play a vital role
in testing Yukawa unified models.

In the case of spin dependent cross section, we show in Fig.~\ref{SIneuSDneu} the
current bounds from the Super-K \cite{Desai:2004pq} and IceCube
\cite{Abbasi:2009uz} experiments and the projected reach of IceCube
DeepCore. It should be noted that IceCube currently
is sensitive only to relatively large neutralino masses and therefore does
not constrain the parameter space that we have
considered. Likewise, while Super-K is sensitive in this region, the
bounds are not stringent enough to rule out anything. However,
from Fig.~\ref{SIneuSDneu} we see that the future IceCube DeepCore
experiment will be able to constrain a significant region of the parameter space.

\begin{figure}[h!]
\centering

\includegraphics[width=8cm]{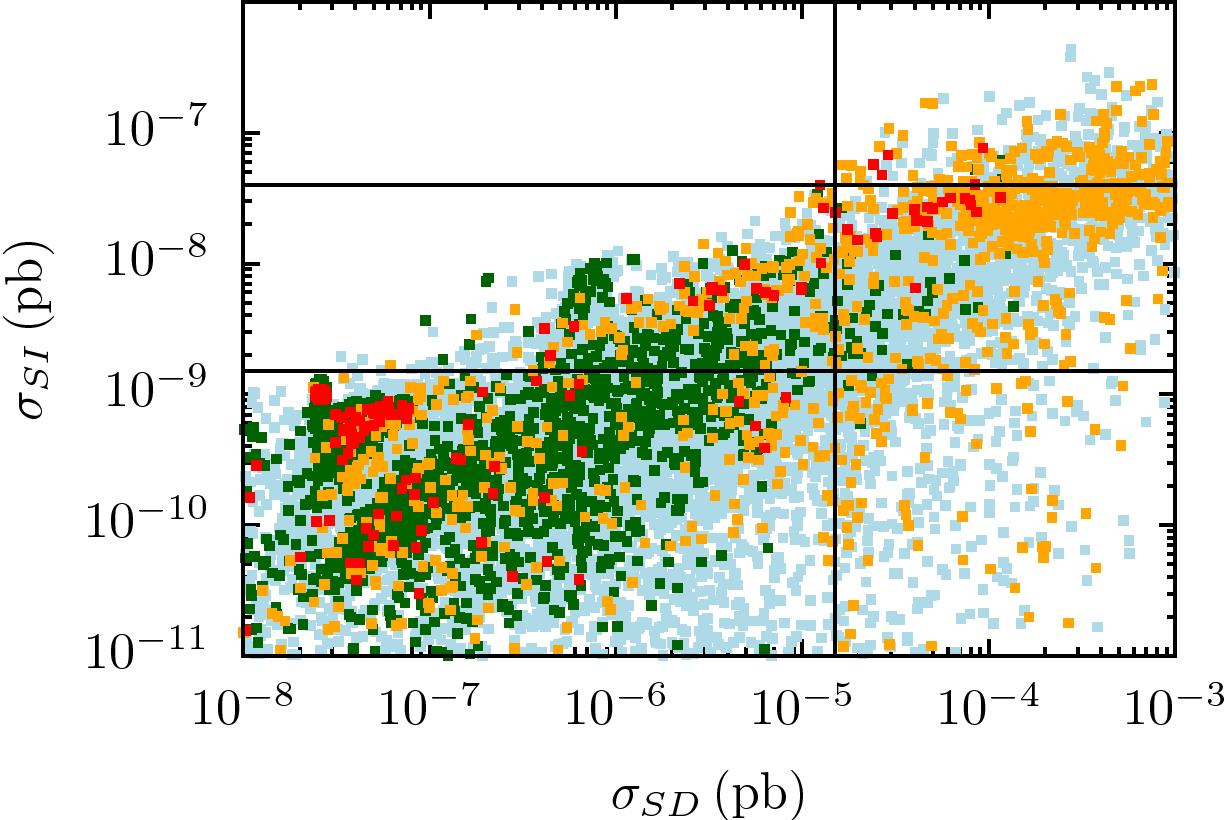}

\caption{Correlation of $\sigma_{\rm SI}$
and $\sigma_{\rm SD}$. Color coding
same as in Fig.~\ref{spectra}. Also shown are lines corresponding to
$\sigma_{\rm SI}=4\times 10^{-8},1.5\times 10^{-9} {\rm(pb)}$
and $\sigma_{\rm SD}=1.5\times10^{-5}{\rm (pb)}$.
\label{SISD}
}
\end{figure}

It is interesting to comment on the correlation between spin independent
and spin dependent cross sections. While both of these
cross sections are enhanced by the presence of a larger higgsino
component in the neutralino, the spin independent cross section
falls off as $1/m_A^4$. In Fig.~\ref{SISD} we plot the correlation
of the spin independent cross section versus the spin dependent
cross section for $m_{\tilde{ \chi}_1^0}\gtrsim 50\, {\rm GeV}$. We
impose this lower bound on the neutralino mass only because for
$50\,  {\rm GeV}< m_{\tilde{ \chi}_1^0}< 400\, {\rm GeV}$, both the
spin independent and spin dependent future reach and spin
independent current bounds are nearly flat. This provides one with an
opportunity to identify regions with simultaneously high spin
independent and spin dependent cross section which may have
the best hope for being tested in ongoing and future
experiments.

In Fig.~\ref{SISD} we
also show lines corresponding to $\sigma_{\rm SI}=4\times
10^{-8},1.5\times 10^{-9} {\rm(pb)}$ and $\sigma_{\rm
SD}=1.5\times10^{-5}{\rm (pb)}$. These lines serve as a guide to
demonstrate the cross correlation of spin independent and spin
dependent cross sections and the plausibility of either discovering
a Yukawa unified neutralino dark matter solution or definitively
ruling out certain regions of the allowed parameter space. It is to
be noted that since ongoing and future experiments require large
spin independent and spin dependent cross sections,
neutralino coannihilation channels such as stau coannihilation,
bino-wino coannihilation and gluino coannihilation may be
difficult to test with direct as well as indirect
detection experiments such as IceCube
DeepCore.

Let us remark on the low mass neutralinos
that we have found in this model.
Because of the relative negative sign between $M_2$ and $M_3$, it is
possible in principle to have $M_1\sim 0$. This means that within
this framework, the neutralino may be as light as we want. The
neutralino mass nonetheless is bounded from below because of the
relic density bounds on dark matter. The 4-2-2 model as we have
presented it has all the ingredients needed to lower the neutralino
mass to the lowest possible value allowed by various constraints. We do
not focus on finding the lightest neutralino in this model. Of the
data that we collected, the lightest neutralino found that is
consistent with Yukawa unification has mass $\sim 43\, {\rm GeV}$. If
we do not insist on Yukawa unification, we can get a neutralino as light as
$\sim 32\, {\rm GeV}$. If we are willing to give up neutralino dark matter, then
$M_1\sim 0$ consistent with Yukawa unification is possible as is evident
in Fig.~\ref{spectra}. This, of course, requires invoking some other
dark matter candidate such as axino.

Finally in Table~\ref{table1} we present some benchmark points for the 4-2-2 Yukawa unified
model with $\mu<0$. All of these points are
consistent with neutralino dark matter and the constraints
mentioned in Section~\ref{constraintsSection}. Point 1 corresponds
to a solution with perfect Yukawa unification ($R=1.0$). Point 2
represents a solution with minimum neutralino mass ($43\, {\rm GeV}$)
consistent with Yukawa unification. Point 3 has a significant
bino-higgsino admixture and, therefore, has relatively large spin independent
and spin dependent neutralino-proton scattering cross-sections. Point
4 depicts a solution with a stop mass of only $826\, {\rm GeV}$. Finally
in point 5 we show an example with bino-gluino coannihilation channel
with the gluino as light as $259 \, {\rm GeV}$. This should be relatively
easy to find at the LHC.

\section{Conclusions\label{conclusions}}

We have shown that Yukawa coupling unification consistent with known
experimental constraints is realized in a SUSY $SU(4)_c \times
SU(2)_L \times SU(2)_R$ model. With $\mu<0$ Yukawa coupling
unification is achieved for  $m_0 \gtrsim 400\, {\rm GeV}$, as opposed
to $m_0 \gtrsim 8\, {\rm TeV}$ for $\mu>0$, by taming the finite corrections to
the b-quark mass. By considering $M_2 <0$ and $M_3>0$ gauginos
and $\mu<0$, we can obtain the correct sign for the desired contribution to $(g-2)_\mu$.
This enables us to simultaneously satisfy the requirements of  $t-b-\tau$ Yukawa
unification, neutralino dark matter and $(g-2)_\mu$, as well as a
variety of other known bounds. We have demonstrated the existence
of a variety of coannihilation scenarios involving gluino, wino
and stau, in addition to the light Higgs, Z and $A$ resonance
solutions. The Yukawa unified solutions may also have
relativley large spin independent and spin dependent interaction cross sections
with nucleons in the case of mixed bino-Higgsino dark matter.
Finally, within the 4-2-2 model, it
is possible to obtain relatively low neutralino masses $\sim 43 \, {\rm
GeV}$ ($\sim 30 \,{\rm GeV}$ without Yukawa unification) consistent with neutralino dark matter.

\begin{table}[h]
\centering
\begin{tabular}{lccccc}
\hline
\hline
                 & Point 1 & Point 2    & Point 3 & Point 4 & Point 5   \\
\hline
$m_{0}$          &  1027       & 1800    & 1210    & 980    & 1720            \\
$M_{1} $         &  -665       & -81     & -414    & -126   & -538             \\
$M_{2} $         &  -1475      & -543    & -940    & -517   & -943             \\
$M_{3} $         &  550       & 611     & 374     & 460    & 70                \\
$\tan\beta$      &  49.1      & 52.8    & 50.6    & 47.0   & 47.6              \\
$A_0/m_0$        &  0.26      & 1.06    & -1.15   & -1.08  & -1.25            \\
$m_{Hu}$         &  743       & 1919    & 1231    & 1090   & 295              \\
$m_{Hd}$         &  1505       & 2395    & 1745    & 1869   & 1729              \\

\hline
$m_h$            &  114       & 115     & 114     & 115   & 115     \\
$m_H$            &  847       & 573     & 781     & 1100  & 1006      \\
$m_A$            &  841       & 569     & 776     & 1090  & 1000      \\
$m_{H^{\pm}}$    &  852       & 581     & 787     & 1100  & 1010     \\

\hline
$m_{\tilde{\chi}^0_{1,2}}$
                 &  280,341    & 43,352  & 168,242 & 56,337 & 233,782     \\
$m_{\tilde{\chi}^0_{3,4}}$
                 &  352,1236 & 380,513 & 246,795 & 371,476& 1210,1216 \\

$m_{\tilde{\chi}^{\pm}_{1,2}}$
                 & 342,1225  & 355,509  & 239,786 & 338,475& 782,1217 \\
$m_{\tilde{g}}$  & 1321    &   1470     & 955     & 1110   & 270  \\

\hline $m_{ \tilde{u}_{L,R}}$
                 & 1771,1489  & 2170,2130  & 1550,1410& 1400,1320& 1818,1697    \\
$m_{\tilde{t}_{1,2}}$
                 & 1053,1410  & 1400,1440  & 822,1040 & 826,965  & 1070,1248  \\
\hline $m_{ \tilde{d}_{L,R}}$
                 & 1773,1512 & 2180,2160   & 1550,1440& 1400,1370& 1820,1730  \\
$m_{\tilde{b}_{1,2}}$
                 & 954,1399 & 1350,1430   & 774,1020  &724,906  & 992,1245  \\
\hline
$m_{\tilde{\nu}_{1}}$
                 & 1391       &  1810     &  1340      & 1000     & 1807  \\
$m_{\tilde{\nu}_{3}}$
                 & 1211       &  1420     &  1100      & 759      & 1550  \\
\hline
$m_{ \tilde{e}_{L,R}}$
                &  1393,1096   &  1820,1820 & 1340,1250 &1010,1040  & 1809,1763  \\
$m_{\tilde{\tau}_{1,2}}$
                &  500,1212   &  885,1420  & 641,1110  &462,765  &  1170,1554 \\
\hline

$\sigma_{SI}({\rm pb})$
                & $4.02\times 10^{-8}$ & $4.1\times 10^{-9}$ & $4.1\times 10^{-8}$
                & $9.5\times 10^{-10}$ & $1.1\times 10^{-10}$      \\

$\sigma_{SD}({\rm pb})$
                & $8.4 \times 10^{-5}$ & $7.5 \times 10^{-6}$ & $1.7\times 10^{-4}$
                & $8.2\times 10^{-6}$ & $2.9\times 10^{-8}$   \\

$\Omega_{CDM}h^2$
                & 0.08       & 0.11   &  0.09     & 0.08    & 0.11 \\

$R$             & 1.01        & 1.11   &  1.09     & 1.07    & 1.08 \\
$g_3$/$g_1 (M_{\rm GUT})$     & 0.98       & 0.98   &  0.99     & 0.98    & 1.00 \\
\hline
\hline
\end{tabular}
\caption{ Sparticle and Higgs masses (in GeV),
with $m_t=173.1\, {\rm GeV}$. All of these benchmark points satisfy the various constraints
mentioned in Section~\ref{constraintsSection} and are compatible with Yukawa unification.
Point 1 exhibits `perfect' Yukawa unification, point 2 has the lightest neutralino, point 3
shows `large' spin independent and spin dependent cross-sections, points 4 and 5
correspond to the lightest stop and gluino respectively. Point 5 also provides an
example of bino-gluino coannihilation channel.
\label{table1}}
\end{table}

\acknowledgments

We thank Howie Baer and Azar Mustafayev for valuable discussions. This work
is supported in part by the DOE Grant No. DE-FG02-91ER40626
(I.G., R.K., S.R. and Q.S.) and GNSF Grant No. 07\_462\_4-270 (I.G.).


\begin{thebibliography}{99}

\bibitem{big-422}
B. Ananthanarayan, G. Lazarides and Q. Shafi, Phys. Rev. D {\bf 44},
1613 (1991) and Phys. Lett. B {\bf 300}, 24 (1993)5; Q.~Shafi and
B.~Ananthanarayan, Trieste HEP Cosmol.1991:233-244; L.~J.~Hall,
R.~Rattazzi and U.~Sarid, Phys.\ Rev.\  D {\bf 50}, 7048 (1994).



\bibitem{bigger-422}
V. Barger, M. Berger and P. Ohmann, Phys. Rev. D {\bf 49}, (1994)
4908; M. Carena, M. Olechowski, S. Pokorski and C. Wagner,  Nucl.\
Phys.\  B {\bf 426}, 269 (1994); B. Ananthanarayan, Q. Shafi and X.
Wang, Phys. Rev. D {\bf 50}, 5980 (1994); G. Anderson et al. Phys.
Rev. D {\bf 47}, (1993) 3702 and Phys. Rev. D {\bf 49},  3660
(1994); R. Rattazzi and U. Sarid, Phys. Rev. D {\bf 53}, 1553
(1996); T. Blazek, M. Carena, S. Raby and C. Wagner, Phys. Rev. D
{\bf 56}, 6919 (1997); T. Blazek, S. Raby and K. Tobe, Phys. Rev. D
{\bf 62}, 055001 (2000); H. Baer, M. Diaz, J. Ferrandis and X. Tata,
Phys. Rev. D {\bf 61}, 111701 (2000); H. Baer, M. Brhlik, M. Diaz,
J. Ferrandis, P. Mercadante, P. Quintana and X. Tata, Phys. Rev. D
{\bf 63}, 015007(2001); S. Profumo, Phys. Rev. D {\bf 68} (2003)
015006; C.~Balazs and R.~Dermisek, JHEP {\bf 0306}, 024 (2003);
C. Pallis, Nucl. Phys. B {\bf 678},  398 (2004); M. Gomez,
G. Lazarides and C. Pallis, Phys. Rev. D {\bf 61} (2000) 123512,
Nucl. Phys. B {\bf 638},  165 (2002) and Phys. Rev. D {\bf 67},
097701(2003);  I.~Gogoladze, Y.~Mimura, S.~Nandi and K.~Tobe, Phys.\
Lett.\  B {\bf 575}, 66 (2003); U. Chattopadhyay, A. Corsetti and P.
Nath, Phys. Rev. D {\bf 66} 035003, (2002); T.~Blazek, R.~Dermisek
and S.~Raby, Phys.\ Rev.\ Lett.\  {\bf 88}, 111804 (2002) and Phys.\
Rev.\  D {\bf 65}, 115004 (2002); M. Gomez, T. Ibrahim, P. Nath and
S. Skadhauge, Phys. Rev. D {\bf 72}, 095008 (2005); K. Tobe and J.
D. Wells, Nucl. Phys. B {\bf 663}, 123 (2003); W.~Altmannshofer,
D.~Guadagnoli, S.~Raby and D.~M.~Straub, Phys.\ Lett.\  B {\bf 668},
385 (2008); D.~Guadagnoli, S.~Raby and D.~M.~Straub, JHEP {\bf 0910}, 059 (2009);
 H.~Baer, S.~Kraml and S.~Sekmen, JHEP {\bf 0909}, 005 (2009);
  K.~Choi, D.~Guadagnoli, S.~H.~Im and C.~B.~Park,
  arXiv:1005.0618 [hep-ph].


\bibitem{Baer:2008jn}
  H.~Baer, S.~Kraml, S.~Sekmen and H.~Summy,
  JHEP {\bf 0803}, 056 (2008);
 H.~Baer, M.~Haider, S.~Kraml, S.~Sekmen and H.~Summy,
  JCAP {\bf 0902}, 002 (2009).



\bibitem{Gogoladze:2009ug}
  I.~Gogoladze, R.~Khalid and Q.~Shafi,
  Phys.\ Rev.\  D {\bf 79}, 115004 (2009).

\bibitem{Baer:2009ff}
 H.~Baer, S.~Kraml, A.~Lessa and S.~Sekmen,
 JHEP {\bf 1002}, 055 (2010)
 [arXiv:0911.4739 [hep-ph]].




\bibitem{Gogoladze:2009bn}
  I.~Gogoladze, R.~Khalid and Q.~Shafi,
  Phys.\ Rev.\  D {\bf 80}, 095016 (2009).

\bibitem{pati}
J.~C.~Pati and A.~Salam,
  Phys.\ Rev.\  D {\bf 10}, 275 (1974).

\bibitem{lr}
  R.~N.~Mohapatra and J.~C.~Pati,
  Phys.\ Rev.\  D {\bf 11}, 2558 (1975);
   G.~Senjanovic and R.~N.~Mohapatra,
  Phys.\ Rev.\  D {\bf 12}, 1502 (1975);
  M.~Magg, Q.~Shafi and C.~Wetterich,
  Phys.\ Lett.\  B {\bf 87}, 227 (1979);
    M.~Cvetic,
  Nucl.\ Phys.\  B {\bf 233}, 387 (1984).


\bibitem{c-parity}
  T.~W.~B.~Kibble, G.~Lazarides and Q.~Shafi,
  Phys.\ Lett.\  B {\bf 113}, 237 (1982);
  T.~W.~B.~Kibble, G.~Lazarides and Q.~Shafi,
  Phys.\ Rev.\  D {\bf 26}, 435 (1982);
  R.~N.~Mohapatra and B.~Sakita,
  Phys.\ Rev.\  D {\bf 21}, 1062 (1980).

\bibitem{Profumo:2004wk}
  S.~Profumo and C.~E.~Yaguna,
  Phys.\ Rev.\  D {\bf 69}, 115009 (2004);
  D.~Feldman, Z.~Liu and P.~Nath,
  Phys.\ Rev.\  D {\bf 80}, 015007 (2009).

\bibitem{Bennett:2006fi}
  G.~W.~Bennett {\it et al.}  [Muon G-2 Collaboration],
  Phys.\ Rev.\  D {\bf 73}, 072003 (2006).

\bibitem{Stockinger:2006zn}
 For review see for instance D.~Stockinger,
  J.\ Phys.\ G {\bf 34}, R45 (2007)
  [arXiv:hep-ph/0609168].

\bibitem{Chamseddine:1982jx}
  A.~H.~Chamseddine, R.~L.~Arnowitt and P.~Nath,
  Phys.\ Rev.\ Lett.\  {\bf 49}, 970 (1982).
R.~Barbieri, S.~Ferrara, and C.~A. Savoy,
{ Phys. Lett. B} {\bf 119},  343 (1982);
 L.~J. Hall, J.~D. Lykken, and S.~Weinberg,
  {Phys. Rev. D} {\bf 27},  2359 (1983);
E.~Cremmer, P.~Fayet, and L.~Girardello,
 { Phys. Lett. B} {\bf 122},   41 (1983);
N.~Ohta,
{ Prog.
  Theor. Phys.} {\bf 70},  542 (1983).





\bibitem{Hebecker:2001jb}
 See, for instance
  A.~Hebecker and J.~March-Russell,
  Nucl.\ Phys.\  B {\bf 625}, 128 (2002).




\bibitem{su8}
I.~Gogoladze, Y.~Mimura and S.~Nandi, Phys.\ Lett.\  B {\bf 562},
307 (2003);
  Phys.\ Rev.\  D {\bf 69}, 075006 (2004);
  I.~Gogoladze, C.~A.~Lee, Y.~Mimura and Q.~Shafi,
  Phys.\ Lett.\  B {\bf 649}, 212 (2007).



\bibitem{ISAJET}
H.~Baer, F.~E.~Paige, S.~D.~Protopopescu and X.~Tata,
arXiv:hep-ph/0001086.



\bibitem{Hisano:1992jj}
J.~Hisano, H.~Murayama, and T.~Yanagida,
  { Nucl. Phys.} {\bf B402} (1993) 46.
Y.~Yamada,
{ Z. Phys.} {\bf C60} (1993) 83;
 J.~L.~Chkareuli and I.~G.~Gogoladze,
  Phys.\ Rev.\  D {\bf 58}, 055011 (1998).




\bibitem{Gomez:2009yc}
  For recent discussions and additional references see
  V.~Barger, D.~Marfatia and A.~Mustafayev,
  Phys.\ Lett.\  B {\bf 665}, 242 (2008);
  M.~E.~Gomez, S.~Lola, P.~Naranjo and J.~Rodriguez-Quintero,
  arXiv:0901.4013 [hep-ph].



\bibitem{Pierce:1996zz}
D.~M. Pierce, J.~A. Bagger, K.~T. Matchev, and R.-j. Zhang,
  { Nucl. Phys.} {\bf B491} (1997) 3.

\bibitem{Ibanez:1982fr}
L.~E. Ibanez and G.~G. Ross,
 { Phys. Lett.} {\bf B110} (1982) 215;
K.~Inoue, A.~Kakuto, H.~Komatsu and S.~Takeshita,
 { Prog. Theor. Phys.} {\bf 68}, 927 (1982)
 [Erratum-ibid.\  {\bf 70}, 330 (1983)];
L.~E. Ibanez,
{ Phys.  Lett.} {\bf B118} (1982) 73;
 J.~R. Ellis, D.~V. Nanopoulos,
and K.~Tamvakis,
  { Phys. Lett.} {\bf B121} (1983) 123;
L.~Alvarez-Gaume, J.~Polchinski, and M.~B. Wise,
{ Nucl. Phys.} {\bf B221} (1983) 495.









\bibitem{PDG}
  C.~Amsler {\it et al.}  [Particle Data Group],
  Phys.\ Lett.\  B {\bf 667}, 1 (2008).





\bibitem{:2009ec}
    [Tevatron Electroweak Working Group and CDF Collaboration and D0 Collab],
  arXiv:0903.2503 [hep-ex].


\bibitem{Belanger:2009ti}
  G.~Belanger, F.~Boudjema, A.~Pukhov and R.~K.~Singh,
  JHEP {\bf 0911}, 026 (2009);
H.~Baer, S.~Kraml, S.~Sekmen and H.~Summy,
  JHEP {\bf 0803}, 056 (2008).



\bibitem{Amsler:2008zzb}
  C.~Amsler {\it et al.}  [Particle Data Group],
  Phys.\ Lett.\  B {\bf 667}, 1 (2008).


\bibitem{Baer:2002fv}
H.~Baer, C.~Balazs, and A.~Belyaev,
   { JHEP} {\bf 03} (2002) 042;
 H.~Baer, C.~Balazs, J.~Ferrandis, and X.~Tata
  { Phys. Rev.} {\bf D64} (2001)  035004.





\bibitem{Schael:2006cr}
  S.~Schael {\it et al.}  
  Eur.\ Phys.\ J.\  C {\bf 47}, 547 (2006).


\bibitem{:2007kv}
  T.~Aaltonen {\it et al.}  [CDF Collaboration],
  Phys.\ Rev.\ Lett.\  {\bf 100}, 101802 (2008).


\bibitem{Barberio:2008fa}
  E.~Barberio {\it et al.}  [Heavy Flavor Averaging Group],
  arXiv:0808.1297 [hep-ex].


\bibitem{Komatsu:2008hk}
  E.~Komatsu {\it et al.}  [WMAP Collaboration],
  Astrophys.\ J.\ Suppl.\  {\bf 180}, 330 (2009).







\bibitem{Ahmed:2009zw}
  Z.~Ahmed {\it et al.}  [The CDMS-II Collaboration],
  arXiv:0912.3592 [astro-ph.CO].



\bibitem{Aprile:2010um}
  E.~Aprile {\it et al.}  [XENON100 Collaboration],
  arXiv:1005.0380 [astro-ph.CO].

\bibitem{Desai:2004pq}
  S.~Desai {\it et al.}  [Super-Kamiokande Collaboration],
  Phys.\ Rev.\  D {\bf 70}, 083523 (2004)
  [Erratum-ibid.\  D {\bf 70}, 109901 (2004)].


\bibitem{Abbasi:2009uz}
  R.~Abbasi {\it et al.}  [ICECUBE Collaboration],
  Phys.\ Rev.\ Lett.\  {\bf 102}, 201302 (2009).











\end{thebibliography}
\end{document}